\documentclass[aps,prx,superscriptaddress,amsfonts,amsmath,amssymb,reprint,showpacs,floatfix]{revtex4-1}

\usepackage{url}
\usepackage{bm}
\usepackage{graphicx}
\usepackage{amsmath}
\usepackage{amstext}
\usepackage{amssymb}
\usepackage{amsfonts}
\usepackage{amsbsy}
\usepackage{verbatim}
\usepackage{color}
\usepackage[colorlinks=true, urlcolor=blue, linkcolor=blue, citecolor=blue, pdftex]{hyperref}
\usepackage{multirow}
\usepackage{pdfpages}
\usepackage{floatrow}
\usepackage{gensymb}
\usepackage{textcomp}

\begin{document}

\title{Spin liquid nature in the Heisenberg $J_{1}-J_{2}$ triangular antiferromagnet}

\author{Yasir Iqbal}
\email[]{yiqbal@physik.uni-wuerzburg.de}
\affiliation{Institute for Theoretical Physics and Astrophysics, Julius-Maximilian's University of W\"urzburg, Am Hubland, D-97074 W\"urzburg, Germany}
\author{Wen-Jun Hu}
\email[]{wenjun.hu@rice.edu}
\affiliation{Department of Physics and Astronomy, Rice University, Houston, Texas 77005-1827, USA}
\author{Ronny Thomale}
\email[]{rthomale@physik.uni-wuerzburg.de}
\affiliation{Institute for Theoretical Physics and Astrophysics, Julius-Maximilian's University of W\"urzburg, Am Hubland, D-97074 W\"urzburg, Germany}
\author{Didier Poilblanc}
\email[]{didier.poilblanc@irsamc.ups-tlse.fr}
\affiliation{Laboratoire de Physique Th\'eorique UMR-5152, CNRS and Universit\'e de Toulouse, F-31062 Toulouse, France}
\author{Federico Becca}
\email[]{becca@sissa.it}
\affiliation{Democritos National Simulation Center, Istituto Officina dei Materiali del CNR and 
SISSA-International School for Advanced Studies, Via Bonomea 265, I-34136 Trieste, Italy}

\date{\today}

\begin{abstract}

We investigate the spin-$\frac{1}{2}$ Heisenberg model on the triangular lattice in the presence of nearest-neighbor $J_1$ and
next-nearest-neighbor $J_2$ antiferromagnetic couplings. Motivated by recent findings from density-matrix renormalization group 
(DMRG) claiming the existence of a gapped spin liquid with signatures of spontaneously broken lattice point group symmetry 
[Zhu and White, \href{http://dx.doi.org/10.1103/PhysRevB.92.041105}{Phys. Rev. B {\bf 92}, 041105 (2015)}; Hu, Gong, Zhu, and 
Sheng, \href{http://dx.doi.org/10.1103/PhysRevB.92.140403}{Phys. Rev. B {\bf 92}, 140403 (2015)}], we employ the variational 
Monte Carlo (VMC) approach to analyze the model from an alternative perspective that considers both magnetically ordered and 
paramagnetic trial states. We find a quantum paramagnet in the regime $0.08\lesssim J_2/J_1\lesssim 0.16$, framed by $120\degree$
coplanar (stripe collinear) antiferromagnetic order for smaller (larger) $J_2/J_1$. By considering the optimization of 
spin-liquid wave functions of a different gauge group and lattice point group content as derived from Abrikosov mean-field theory, 
we obtain the gapless $U(1)$ Dirac spin liquid as the energetically most preferable state in comparison to all symmetric or 
nematic gapped $\mathbb{Z}_{2}$ spin liquids so far advocated by DMRG. Moreover, by the application of few Lanczos iterations,
we find the energy to be the same as the DMRG result within error-bars. To further resolve the intriguing disagreement between 
VMC and DMRG, we complement our methodological approach by the pseudofermion functional renormalization group (PFFRG) to compare 
the spin structure factors for the paramagnetic regime calculated by VMC, DMRG, and PFFRG. This model promises to be an ideal 
test-bed for future numerical refinements in tracking the long-range correlations in frustrated magnets.
\end{abstract}


\maketitle

\section{Introduction}\label{sec:intro}

Quantum antiferromagnetic models on two-dimensional frustrated lattices provide a natural habitat for the birth of novel quantum 
spin-liquid states~\cite{Pomeranchuk-1941,Anderson-1973,Fazekas-1974}, whose search has been a keynote of contemporary condensed 
matter physics~\cite{Balents-2010}. For example, spin-$\frac{1}{2}$ Heisenberg models defined on the kagome lattice have been 
shown to potentially host exotic spin liquids, sometimes with controversial findings from different numerical methods. 
This includes potential microscopic models for the chiral spin liquid as originally described by Kalmeyer and Laughlin and 
similar states~\cite{Kalmeyer-1987,Schroeter-2007,Thomale-2009,Greiter-2014,Nielsen-2012,He-2014,Sheng-2014,Bauer-2014,Gong-2015,He-2015,Zhu-2015,Hu-2015a,Kumar-2015,Wietek-2015,Pollmann-2015,Fuji-2015}, 
the gapped (topological) $\mathbb{Z}_2$ spin liquid proposed to describe the properties of the nearest-neighbor model of this 
highly frustrated lattice~\cite{Wen-1991,Yan-2011,Depenbrock-2012,Jiang-2012}, the foundation of paradigmatic gapless spin 
liquids such as the $U(1)$ Dirac spin liquid and algebraic spin liquids~\cite{Ran-2007,Hermele-2008,Iqbal-2013,Hu-2015b}, and 
attempts to resolve magnetic phase diagrams assisting the experimental investigation of Herbertsmithite crystals and polymorphs 
thereof~\cite{Suttner-2014,Kolley-2015,Iqbal-2015a,Bieri-2015,Iqbal-2015c}. 

Another prominent candidate model conjectured to host a quantum paramagnetic ground state is the spin-$\frac{1}{2}$ triangular 
lattice with both antiferromagnetic nearest- ($J_1$) and next-nearest-neighbor ($J_2$) couplings~\cite{Baskaran-1989,Jolicoeur-1990}. 
Although there are several compounds in which magnetic moments lie on stacked layers with a triangular geometry, most of them have 
sizable distortions, leading to spatial anisotropies along different directions~\cite{Coldea-2002,Zvyagin-2014}. Very recently, it 
has been claimed that Ba$_3$CoSb$_2$O$_9$ gives an almost perfect realization of a spin-$\frac{1}{2}$ equilateral triangular lattice 
antiferromagnet, with both $J_1$ and $J_2$ couplings~\cite{Ma-2015}. From a theoretical point of view, the classical limit of the 
$J_1{-}J_2$ model has three different phases: for $J_2/J_1<1/8$, the system has three-sublattice $120\degree$ coplanar order, for 
$1/8<J_2/J_1<1$ it is infinitely degenerate (with four-sublattice periodicity, in which the only constraint is to have the four spins 
sum to zero), and for $J_2/J_1>1$ it features generic incommensurate spiral structures. By including spin-wave fluctuations, both at 
the lowest (first) and second orders, the coplanar phase remains stable, while the accidental degeneracy of the intermediate phase is 
lifted in favor of a stripe collinear order with two-sublattice periodicity~\cite{Jolicoeur-1990,Chubukov-1992}. Naturally, quantum 
paramagnetic domains tend to emerge in the vicinity of classical transition points, i.e., $J_2/J_1=1/8$ and $J_2/J_1=1$; however, 
their actual stabilization is not clear within spin-wave approaches~\cite{Jolicoeur-1990,Chubukov-1992}. Subsequent works have shown 
conflicting results on the possible existence, extent, and nature of nonmagnetic phases~\cite{Chubukov-1992,Korshunov-1993,Ivanov-1993,Deutscher-1993,Lecheminant-1995,Sindzingre-1995,Manuel-1999,Li-2015}. 
Some more recent studies have vouched for the existence of a quantum paramagnet in the vicinity of $J_2/J_1=1/8$, while the 
problem of the precise identification of its nature and extent in parameter space remains an open issue: a Schwinger-boson 
approach found the corresponding window to be $0.12\lesssim J_2/J_1 \lesssim 0.19$ with no further clarification of the nature of
the paramagnetic state~\cite{Manuel-1999}, while a high-order coupled-cluster method (CCM) study predicted a quantum paramagnet 
for $0.060(10) \leqslant J_2/J_1 \leqslant 0.165(5)$~\cite{Li-2015}, with a spin-triplet gap which vanishes in the entire paramagnetic
regime~\cite{Bishop-2015}. In addition, two different variational Monte Carlo (VMC) studies claimed for a gapless spin liquid
close to $J_2/J_1=1/8$: Kaneko and co-workers~\cite{Kaneko-2014} used a full optimization of the pairing of a Gutzwiller-projected 
BCS wave function [obtaining a critical spin liquid for $0.10(1) \leqslant J_2/J_1 \leqslant 0.135(5)$] and Mishmash and 
collaborators~\cite{Mishmash-2013} considered few variational {\it Ans\"atze} to describe both magnetic and nonmagnetic phases (here, 
they obtained evidence for a gapless nodal $d$-wave spin liquid for $0.06 \lesssim J_2/J_1 \lesssim 0.17$).
In the former case, the full optimization of the pairing function faces technical difficulties, which make it difficult to reach
true energy minima; in the latter one, the variational states are relatively simple and do not exhaust the rich variety of
states that can be obtained within the fermionic representation of Gutzwiller-projected states. Indeed, the variational energies
that we get are much better than those of these two papers, indicating the high accuracy of the present approach.

\begin{figure}
\includegraphics[width=1.0\columnwidth]{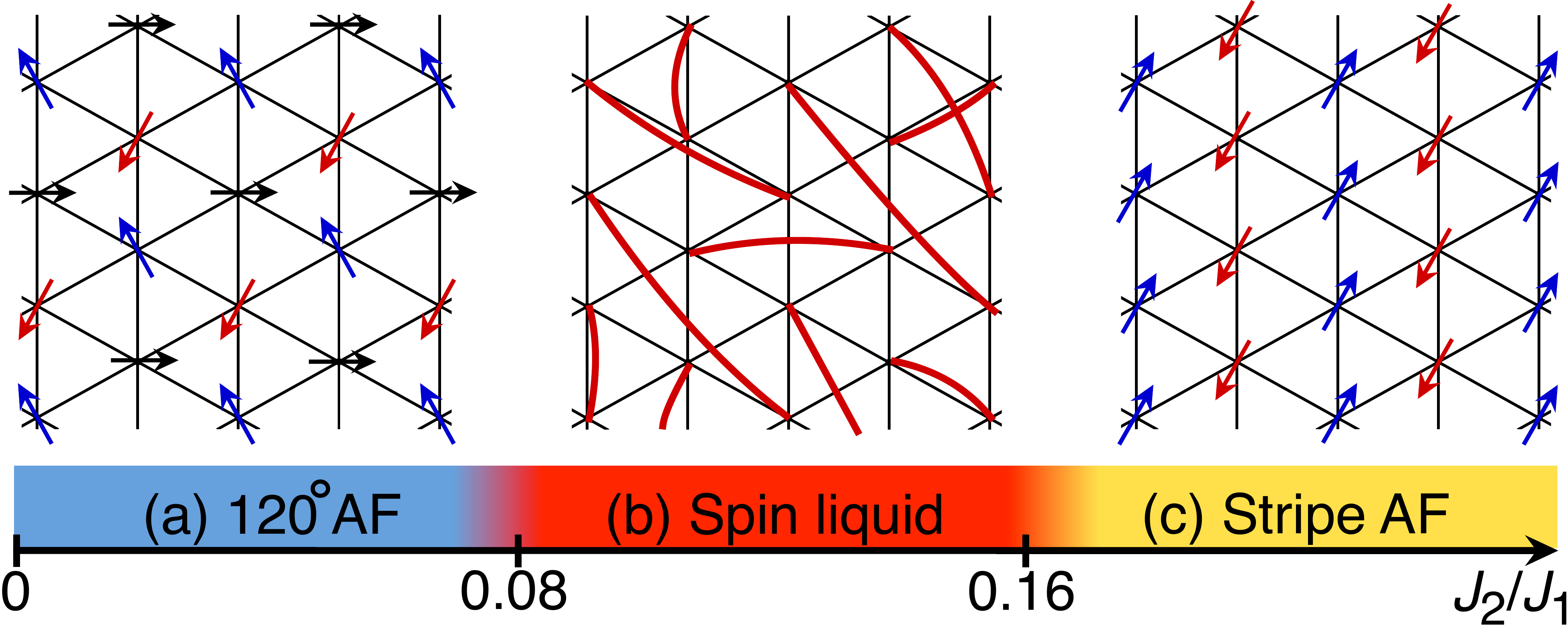}
\caption{\label{fig:fig1}
(Color online) Schematic illustrations of the coplanar three-sublattice (black, blue, and red) magnetic order on the triangular 
lattice (a), the resonating-valence bond spin liquid (b) and the collinear two-sublattice (blue and red) stripe magnetic order 
(c). The phase diagram, as obtained by using variational Monte Carlo method is also reported. Note that the DSL found here 
can be represented as a resonating-valence bond spin liquid with a power-law distribution of bond amplitudes. }
\end{figure}

By contrast, density-matrix renormalization group (DMRG) studies find a gapped $\mathbb{Z}_{2}$ topological spin liquid for 
$0.06 \lesssim J_2/J_1 \lesssim 0.17$~\cite{White-2015} and $0.08 \lesssim J_2/J_1 \lesssim0.16$~\cite{Hu-2015}, with signatures 
of possible spontaneously broken rotational symmetry. Following this proposal, Zheng, Mei, and Qi,~\cite{Zheng-2015}, and, in an 
independent work, Lu~\cite{Lu-2015} have performed a classification of symmetric and nematic $\mathbb{Z}_{2}$ spin liquids, and 
pointed out promising candidates at the fermionic mean-field level. This approach was extended by Bieri, Lhuillier, and 
Messio~\cite{Bieri-2015b} to include chiral spin liquids as well. A bosonic mean-field classification has likewise been 
accomplished~\cite{Wang-2006,Messio-2013}, with some of the states addressed already in earlier 
works~\cite{Sachdev-1992,Misguich-2012}.

In this paper, we address the $J_{1}-J_{2}$ Heisenberg model on the triangular lattice from the viewpoint of versatile Gutzwiller 
projected Abrikosov-fermion wave functions (optionally supplemented by Lanczos optimization), which we implemented by using 
efficient VMC techniques. To enable a comparison of the variational energies, we also perform DMRG and Lanczos diagonalizations for 
specific regions in parameter space. In order to resolve the magnetic susceptibility profile in the paramagnetic regime, we employ 
pseudofermion functional renormalization group (PFFRG) calculations, the results of which are then compared with analogous results 
from DMRG and VMC. Our main VMC results are summarized as follows: a spin-liquid phase is stabilized for 
$0.08 \lesssim J_2/J_1 \lesssim 0.16$ (Fig.~\ref{fig:fig1}), in excellent agreement with DMRG~\cite{White-2015,Hu-2015} and
CCM~\cite{Li-2015}. Within the spin-liquid regime, however, we find no signal of stabilization for any of the gapped symmetric 
or nematic $\mathbb{Z}_{2}$ states proposed in Refs.~\onlinecite{Zheng-2015,Lu-2015}. In particular, the gapped $\mathbb{Z}_{2}$ 
spin liquids are found to have higher energies compared to the gapless $U(1)$ Dirac spin liquid (DSL); the gapless 
$\mathbb{Z}_{2}$ spin liquids suffer the same fate. We find that nematic order only onsets simultaneously with collinear
antiferromagnetic order, which is also supported by the analysis of nematic response functions in PFFRG. On performing a couple 
of Lanczos optimization steps on the VMC variational result, followed by a zero-variance extrapolation, we obtain estimates of 
the exact ground-state and $S=2$ excited-state energies on different cluster sizes. Our estimate of the ground-state energy on 
finite-systems is in excellent agreement with exact diagonalization and other numerical methods. In the thermodynamic limit, our 
estimate of the ground-state energy is equal to the one obtained by DMRG, within error-bars. However, in contrast to DMRG results,
which found a finite spin excitation gap, the $S=2$ gap computed in VMC is found to extrapolate to zero (within error-bars) in the 
thermodynamic limit. These findings strongly point to a gapless spin liquid ground state, yielding a clear disagreement with the 
findings by DMRG. 

The article is organized as follows. In Sec.~\ref{sec:model}, we describe the model Hamiltonian and discuss its finite-size 
spectra obtained from exact diagonalization, followed by a description of the pseudofermion spin representation framework. 
In Sec.~\ref{sec:methods}, the variational Monte Carlo method, the associated wave functions, and the pseudofermion functional 
renormalization group method are explained. In Sec.~\ref{sec:results}, we present the results on, the energy optimization of 
competing variational states, spin excitation gap, spin structure factors, followed by a discussion of the findings from different 
methods. Conclusions are given in Sec.~\ref{sec:conclusions}.

\begin{figure}
\includegraphics[width=1.0\columnwidth]{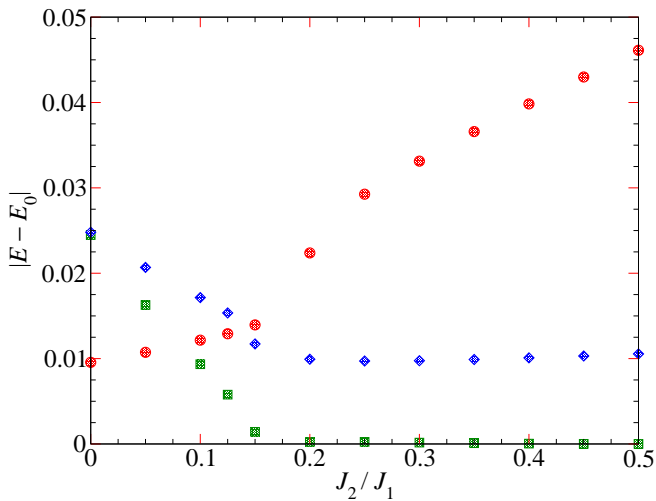}
\caption{\label{fig:fig2}
(Color online) Exact low-energy spectra (measured from the total-symmetric state energy $E_0$) on the $6 \times 6$ cluster for 
different values of $J_2/J_1$. The states are classified according to their spin, momentum $\mathbf{q}$, and eigenvalue $R_{2\pi/6}$ 
of $2\pi/6$ rotations. The total-symmetric state is a singlet with $\mathbf{q}=(0,0)$ and $R_{2\pi/6}=1$. The other states are, (i) 
a singlet with $\mathbf{q}=(0,0)$ and $R_{2\pi/6}=e^{i 2\pi/6}$ (green squares), degenerate with $R_{2\pi/6}=e^{-i 2\pi/6}$, 
(ii) a triplet with $\mathbf{q}=(4\pi/3,0)$ (red circles), which is degenerate with $\mathbf{q}=(2\pi/3,2\pi/\sqrt{3})$, and 
(iii) a triplet with $\mathbf{q}=(0,2\pi/\sqrt{3})$ (blue diamonds), which is degenerate with other two states at symmetry-related 
$\mathbf{q}$ points.}
\end{figure}

\section{Model}\label{sec:model}

The Hamiltonian for the spin-$\frac{1}{2}$ Heisenberg $J_1{-}J_2$ antiferromagnetic model is
\begin{equation}\label{eqn:heis-ham}
{\cal H} = J_1 \sum_{\langle i,j \rangle} \mathbf{S}_{i} \cdot \mathbf{S}_{j}
+J_2 \sum_{\langle\langle i,j \rangle\rangle} \mathbf{S}_{i} \cdot \mathbf{S}_{j},
\end{equation} 
where both $J_1$ and $J_2$ are positive; $\langle i,j \rangle$ and $\langle\langle i,j \rangle\rangle$ denote sums over 
nearest-neighbor (NN) and next-nearest-neighbor (NNN) pairs of sites, respectively. $\mathbf{S}_{i}=(S^x_{i},S^y_{i},S^z_{i})$ 
denotes the spin operator acting on a spin-$\frac{1}{2}$ at site $i$. All energies will be given in units of $J_1$.

\subsection{Finite-size spectra}\label{sec:finite-size-spectra}
Before discussing our main results based on VMC and PFFRG approaches (see Sec.~\ref{sec:results}), we would like to show 
the results of exact diagonalizations on a small $6 \times 6$ cluster. All eigenstates can be classified according to their 
quantum numbers relative to translations (denoted by the momentum $\mathbf{q}$); for particular values of $\mathbf{q}$, also
$\pi/3$ rotations and reflections with respect to the ${\rm x}$-axis can be specified, their quantum numbers being $R_{2\pi/6}$ and
$R_{\rm x}$, respectively. The results for $0 \leqslant J_2/J_1 \leqslant 1/2$ are given in Fig.~\ref{fig:fig2}, where the 
energies per site of few relevant states are reported in comparison with the total-symmetric singlet with $\mathbf{q}=(0,0)$, 
$R_{2\pi/6}=1$ and $R_{\rm x}=1$. In particular, we show (i) a singlet with $\mathbf{q}=(0,0)$ and $R_{2\pi/6}=e^{i 2\pi/6}$,
which is degenerate with the one having $R_{2\pi/6}=e^{-i 2\pi/6}$, (ii) a triplet with $\mathbf{q}=(4\pi/3,0)$, which is 
degenerate with the one having $\mathbf{q}=(2\pi/3,2\pi/\sqrt{3})$, and (iii) a triplet with $\mathbf{q}=(0,2\pi/\sqrt{3})$, 
degenerate with $\mathbf{q}=(\pi,\pm \pi/\sqrt{3})$. The ground state is a total-symmetric singlet for $J_2/J_1 \leqslant 0.15$; 
for $J_2/J_1 \geqslant 0.2$ singlets with $R_{2\pi/6}=e^{\pm i 2\pi/6}$ collapse on it, even having a slightly lower energy 
on this cluster. This fact is compatible with the rise of a phase with stripe collinear order in the thermodynamic limit (or a 
spontaneous breaking of rotational symmetry). In addition, a level crossing appears also in the triplet sector: for small values 
of $J_2/J_1$, the lowest-energy triplet has $\mathbf{q}=(4\pi/3,0)$ (or $\mathbf{q}=(2\pi/3,2\pi/\sqrt{3})$), which is compatible
with the presence of $120\degree$ order in the thermodynamic limit; by contrast for larger values of $J_2/J_1$, the 
lowest-energy triplet has $\mathbf{q}=(0,2\pi/\sqrt{3})$ (or symmetry related momenta), compatible with the collinear magnetic 
order. Unfortunately, on such a small cluster, it is impossible to establish with precision the locations of phase transitions, 
as well as the possible existence of a magnetically disordered phase. Therefore, we address these important questions using 
variational wave function and functional renormalization group approaches, which are described in the ensuing section.

\subsection{Pseudofermion mean-field theory}\label{sec:pseudofermion}
A traditional recipe for the construction of a spin liquid at the mean-field level~\cite{Baskaran-1987,Wen-1991,Wen-2002} rests on 
the introduction of fictitious fermionic fields represented by Abrikosov ``pseudofermion'' operators, $c_{i,\alpha}$ with 
$(\alpha=\uparrow,\downarrow)$, corresponding to spin-$1/2$, charge neutral quasi-particles, called {\it spinons}. 
The physical spin operator $\mathbf{S}_{i}$ can then be expressed as a bilinear in the spinon operators~\cite{Abrikosov-1965}:
\begin{equation}\label{eqn:spinon}
{\mathbf S}_{i}=\frac{1}{2} c_{i,\alpha}^{\dagger}\boldsymbol{\sigma}_{\alpha\beta}c_{i,\beta},
\end{equation}
where the summation over the repeated greek indices is implied and $\boldsymbol{\sigma}=(\sigma^x,\sigma^y,\sigma^z)$ denotes 
the Pauli matrices. This representation is endowed with a {\it local} $SU(2)$ gauge symmetry in 
which~\cite{Baskaran-1988,Affleck-1988,Dagotto-1988}
\begin{equation}
\left(
\begin{array}{c}
c_{i,\uparrow}^{\dagger} \\
c_{i,\downarrow}
\end{array}
\right )
\rightarrow
{\cal U} 
\left(
\begin{array}{c}
c_{i,\uparrow}^{\dagger} \\
c_{i,\downarrow}
\end{array}
\right ),
\end{equation}
where ${\cal U}$ is an $SU(2)$ matrix. In terms of spinon operators the Hamiltonian~(\ref{eqn:heis-ham}) acquires the form
\begin{eqnarray}
&&{\cal H} = \frac{J_1}{2} \sum_{\langle i,j \rangle} \left( c_{i,\alpha}^{\dagger}c_{i,\beta} c_{j,\beta}^{\dagger}c_{j,\alpha}
-\frac{1}{2} c_{i,\alpha}^{\dagger}c_{i,\alpha} c_{j,\beta}^{\dagger}c_{j,\beta} \right) \nonumber \\
&&+ \frac{J_2}{2} \sum_{\langle\langle i,j \rangle\rangle} \left( c_{i,\alpha}^{\dagger}c_{i,\beta} c_{j,\beta}^{\dagger}c_{j,\alpha}
-\frac{1}{2} c_{i,\alpha}^{\dagger}c_{i,\alpha} c_{j,\beta}^{\dagger}c_{j,\beta} \right).
\label{eqn:ff-ham}
\end{eqnarray}
Once the spin operator is written in terms of fermionic operators, the Hilbert space is enlarged. Indeed, within the original
spin model, there are two states per site (i.e., up and down spin), in the fermionic representation, there are four states per site 
(i.e, empty, doubly occupied, and singly occupied with up or down spin). Therefore, in order to describe a legitimate spin state 
in the fermionic representation, one needs to restrict to the sub-space of exactly one-fermion per site with the local constraint:
\begin{equation}\label{eqn:constraint}
c_{i,\alpha}^{\dagger}c_{i,\alpha}=1.
\end{equation} 

Here, we would like to mention that both VMC and PFFRG approaches enforce this constraint {\it exactly}, as described in 
Sec.~\ref{sec:methods}. The PFFRG approach does not assume any particular starting point to treat the Hamiltonian of 
Eq.~(\ref{eqn:ff-ham}), thus avoiding to introduce any possible bias in the calculations; by contrast, within VMC, a 
noninteracting state is considered (and then projected into the correct Hilbert space). In this respect, it is useful to 
briefly discuss the simplest possible approximation of Eq.~(\ref{eqn:ff-ham}), which consists in a mean-field treatment:
\begin{eqnarray}
&&{\cal H}_{{\rm MF}} =
\sum_{(i,j),\alpha}\chi_{ij}c_{i,\alpha}^{\dagger}c_{j,\alpha} + 
\sum_{(i,j)}\Delta_{ij}(c^{\dagger}_{i,\uparrow}c^{\dagger}_{j,\downarrow}+ {\rm H.c.}) \nonumber \\
&&+\sum_{i}\Bigg\{\mu \sum_{\alpha}c_{i,\alpha}^{\dagger}c_{i,\alpha}
+\zeta (c_{i,\uparrow}^{\dagger}c_{i,\downarrow}^{\dagger}+{\rm H.c.})\Bigg\},
\label{eqn:MF-SL}
\end{eqnarray}
where the operators $c_{i,\alpha}^{\dagger}c_{j,\alpha}$ and $c_{i,\uparrow}c_{j,\downarrow}$ have been replaced by their 
corresponding ground-state expectation values, $\chi_{ij}=\langle c_{i,\alpha}^{\dagger}c_{j,\alpha} \rangle$ and 
$\Delta_{ij}=\langle c_{i,\uparrow}c_{j,\downarrow}\rangle$, respectively. The local constraint~(\ref{eqn:constraint}) has been 
replaced by a global one, through the inclusion of a Lagrangian multiplier. Within this approximation, the mean-field ground 
state $|\Psi_{\rm MF}\rangle$ is obtained by diagonalizing Eq.~(\ref{eqn:MF-SL}). However, $|\Psi_{\rm MF}\rangle$ lives in the 
enlarged (i.e., fermionic) Hilbert space and, in order to obtain a legitimate wave function for spins, one must include fluctuations
around the mean-field state. In this respect, an accurate treatment of all (spatial and temporal) fluctuations becomes crucial. 
On a lattice system, it proves impossible to analytically treat all these fluctuations in an accurate manner and one has to resort 
to approximate methods. Temporal fluctuations of the Lagrange multiplier $\mu$ are particularly important, since they enforce the 
one-fermion per-site constraint. In the following, we will describe two possible numerical approaches for describing the ground-state
properties of the spin model. 

\section{Methods}\label{sec:methods}

\subsection{Variational Quantum Monte Carlo}
One possibility to enforce {\it exactly} the one-fermion per site constraint is to apply the Gutzwiller projector to the 
uncorrelated wave function. In this case, a Monte Carlo sampling is needed in order to compute any expectation values over 
variational states, since the resulting wave function includes strong correlations among the fermionic objects. 

Our variational wave functions are defined as:
\begin{equation}\label{eqn:physical-wf}
|\Psi_{\rm var}\rangle = \mathcal{J}_{z}\mathcal{P}_{G}|\Phi_0\rangle.
\end{equation}
Here, $|\Phi_0\rangle$ is an uncorrelated wave function that is obtained as the ground state of a {\it generic} noninteracting 
Hamiltonian, like the one of Eq.~(\ref{eqn:MF-SL}); $\mathcal{P}_{G}=\prod_{i}(n_{i,\uparrow}-n_{i,\downarrow})^2$ is the Gutzwiller 
projector. Notice that $|\Phi_0\rangle$ is obtained without any self-consistent requirement, as in the mean-field approach, but it
is found by minimizing the energy in presence of the Gutzwiller projector. In addition to the Gutzwiller term, a spin-spin 
Jastrow factor is also included to describe magnetically ordered phases:
\begin{equation}\label{eqn:jastrow}
\mathcal{J}_{z}=\exp \left( \frac{1}{2}\sum_{ij}u_{ij}S_{i}^{z}S_{j}^{z} \right ),
\end{equation}
where, $u_{ij}$ is a translationally invariant pseudopotential that depends upon the distance $|{\bf R}_i-{\bf R}_j|$ of two 
sites. All the independent parameters in the pseudopotential are optimized via Monte Carlo simulations. By construction, the 
Jastrow factor breaks the spin $SU(2)$ symmetry of the Heisenberg model. In the following, we consider two cases for the 
noninteracting Hamiltonian that are suitable for generating magnetic and spin-liquid wave functions.

\begin{figure}
\includegraphics[width=1.0\columnwidth]{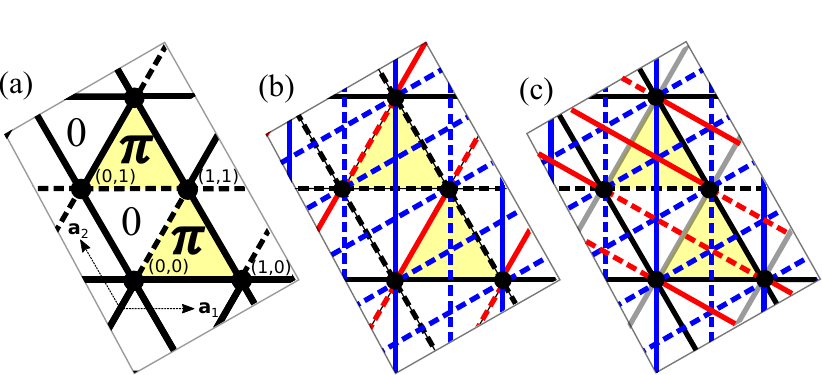}
\caption{\label{fig:fig3}
(Color online) (a) DSL {\it Ansatz}: all NN amplitudes are {\it real} hoppings of equal magnitude. The solid (dashed) bonds denote 
positive (negative) signs. The unit cell is doubled to accommodate the $\pi$ flux. (b) $\mathbb{Z}_{2}\{\pi\}\mathcal{A}$ nematic 
spin-liquid {\it Ansatz} (reproduction of {\it Ansatz} No.~$20$ of Ref.~[\onlinecite{Lu-2015}]): there are two classes of NN hoppings, labeled 
by black and red colors. Non-zero NNN hoppings are denoted by blue color. (c) $\mathbb{Z}_{2}~\mathcal{C}$ nematic spin-liquid 
{\it Ansatz} (reproduction of {\it Ansatz} No.~$6$ of Ref.~[\onlinecite{Lu-2015}]): nonzero NN hoppings are denoted by black color. The grey 
bonds have zero hopping amplitude. There are two classes of NNN hoppings, denoted by blue and red colors. Both these nematic 
{\it Ans\"atze} require a doubling of the unit cell, and allow chemical potential and on-site {\it real} pairing.}
\end{figure}

The magnetic states are defined from:
\begin{equation}\label{eqn:mf-mag}
\mathcal{H}_{\rm MAG}=\sum_{(i,j),\alpha}\chi_{ij}c_{i,\alpha}^{\dagger}c_{j,\alpha} + 
h\sum_{i,\alpha}\mathbf{M}_i\cdot\mathbf{S}_i,
\end{equation}
where $\chi_{ij}$ and $h$ are parameters that can be optimized to minimize the variational energy. The periodicity of the 
magnetic order is defined by the vector $\mathbf{M}_i$. Here, we take $\mathbf{M}_i$ in the $XY$ plane, i.e., 
$\mathbf{M}_i=(\cos(\mathbf{q}\cdot{\bf R}_i),\sin(\mathbf{q}\cdot{\bf R}_i),0)$, where $\mathbf{q}$ is the pitch vector. In 
this way, the Jastrow factor~(\ref{eqn:jastrow}) correctly describes the relevant quantum fluctuations around the classical spin 
state~\cite{Manousakis-1991}. It is worth mentioning that the existence of magnetic long-range order is directly related to the 
presence of a finite parameter $h$. Here, we study two magnetic orders: (i) the three-sublattice $120\degree$ order, 
corresponding to $\mathbf{q}=(4\pi/3,0)$, with the {\it Ansatz} for $\chi_{ij}$ in Eq.~(\ref{eqn:mf-mag}) given by 
Fig.~\ref{fig:fig3}(a) and (ii) a stripe collinear order, corresponding to $\mathbf{q}=(0,2\pi/\sqrt{3})$, with the {\it Ansatz} for 
$\chi_{ij}$ in Eq.~(\ref{eqn:mf-mag}) given by Fig.~\ref{fig:fig3}(b). These choices of $\chi_{ij}$ give the most competitive 
magnetic wave functions.

On the other hand, the spin-liquid wave functions are defined from the noninteracting Hamiltonian of Eq.~(\ref{eqn:MF-SL}), 
where, in addition to the hopping terms there is also a singlet pairing ($\Delta_{ij}=\Delta_{ji}$), chemical potential $\mu$ and 
on-site pairing $\zeta$. Different patterns of distribution of $\chi_{ij}$ and $\Delta_{ij}$, along with a specification of the 
on-site terms $\mu$ and $\zeta$ lead to distinct spin liquids, see Ref.~[\onlinecite{Wen-2002}] for a systematic classification 
scheme. Moreover, the spin-spin Jastrow factor can be also included to improve the variational energy.

When a particle-hole transformation is performed on down electrons: 
\begin{eqnarray}
c^{\dagger}_{i,\downarrow} \to c_{i,\downarrow}, \nonumber \\
c^{\dagger}_{i,\uparrow} \to c^{\dagger}_{i,\uparrow},
\end{eqnarray}
the mean-field Hamiltonian~(\ref{eqn:MF-SL}) commutes with the total number of particles. Therefore the uncorrelated state is 
defined by filling suitable single-particle orbitals. Boundary conditions should be taken in order to have a unique state (i.e., 
filling all orbitals in a shell with the same mean-field energy). Periodic (P) and anti-periodic (A) boundary conditions along 
the ${\bf a}_1$ and ${\bf a}_2$ lattice vectors [see Fig.~\ref{fig:fig3}(a)] can be considered, leading to four choices: [P,P], 
[P,A], [A,P], and [A,A] of boundary conditions.  

The variational parameters in the spin wave function of Eq.~(\ref{eqn:physical-wf}) are optimized using an implementation of the 
stochastic reconfiguration (SR) optimization method~\cite{Sorella-2005,Sorella-2006}. This allows us to obtain an extremely 
accurate determination of variational parameters. Indeed, small energy differences are effectively computed by using a correlated 
sampling, which makes it possible to strongly reduce statistical fluctuations. The current problem of the study of the instability
of gapless spin liquids towards $\mathbb{Z}_{2}$ states will clearly demonstrate the power of this method in navigating 
complicated energy landscapes.

In order to have a systematic improvement of the trial variational wave function and approach the true ground state, we can apply 
a few Lanczos steps to $|\Psi_{\rm var}\rangle$~\cite{Sorella-2001}:
\begin{equation}
\label{eqn:psi-ls}
|\Psi_{p\text{-}\rm{LS}}\rangle =  \bigg{(}1+\sum_{k=1}^{p}\alpha_{k}{\cal H}^k\bigg{)}|\Psi_{\rm var}\rangle.
\end{equation}
Here, the $\alpha_{k}$'s are variational parameters. The convergence of $|\Psi_{p\text{-}\rm{LS}}\rangle$ to the exact ground 
state $|\Psi_{\rm ex}\rangle$ is guaranteed for large $p$ provided the starting state is not orthogonal to 
$|\Psi_{\rm ex}\rangle$, i.e., for $\langle\Psi_{\rm ex}|\Psi_{\rm var}\rangle \neq 0$. On large cluster sizes, only a few steps 
can be efficiently performed and here we implement the case with $p=1$ and $p=2$ ($p=0$ corresponds to the original trial wave 
function). Subsequently, an estimate of the exact ground-state energy may be achieved by the method of variance extrapolation. 
For sufficiently accurate states, we have that $E\approx E_{\rm ex}+{\rm constant}\times\sigma^{2}$, where 
$E=\langle\hat{{\cal H}}\rangle/N$ and $\sigma^{2}=(\langle\hat{{\cal H}}^{2}\rangle-\langle\hat{{\cal H}}\rangle^{2})/N$ 
are the energy and variance per site, respectively, whence, the exact ground-state energy $E_{\rm ex}$ can be extracted by 
fitting $E$ vs $\sigma^{2}$ for $p=0$, $1$, and $2$. The energy and its variance for $p=0$, $1$, and $2$ Lanczos steps are 
obtained using the standard variational Monte Carlo method.

\subsection{Pseudofermion functional renormalization group}
The PFFRG approach~\cite{Reuther-2010,Reuther-2011a} has been successfully employed to different types of hexagonal
magnets~\cite{Reuther-2011a,Reuther-2011b,Reuther-2011c,Singh-2012,Suttner-2014}, and is also based on the decomposition
of the spin operator of Eq.~(\ref{eqn:spinon}), thus requiring to enforce the constraint of one fermion per site. In our 
formalism, the pseudofermion number fluctuations are explicitly not present, since we never allow the system to have any 
diagrammatic contributions that deviate from the one-pseudo-fermion per site constraint. The fermionic representation empowers 
one to exploit Feynman diagrammatic many-body techniques on the Hamiltonian~(\ref{eqn:ff-ham}). Starting from a bare level above 
all spin exchange couplings at half-filling, the bare local free propagator in Matsubara space is given by
$G_{0}(i\omega)=\frac{1}{i\omega}$, wherein the absence of a self-energy term is just a manifestation of the fact that the 
fermionic Hamiltonian~(\ref{eqn:ff-ham}) does not contain any quadratic term. The global Langrange parameter $\mu$ related to 
the single occupancy constraint falls out of the equation because of particle-hole symmetry at half filling. Together, the initial
formulation of the propagator along with the flow equations throughout preserve the strict locality of the propagator in real 
space. The PFFRG framework proceeds by introducing an infrared frequency cutoff $\Lambda$ in the fermion propagator, and by 
triggering a purely imaginary self-energy term induced by the implicit diagrammatic summation, i.e.,
$G_{0}^{\Lambda}(i\omega)=\frac{\Theta(|\omega|-\Lambda)}{i\omega+\Sigma^\Lambda(\omega)}$,
$\lim_{\Lambda \to \infty}\Sigma^\Lambda(\omega)=0$, resulting in a $\Lambda$-dependence of all $m$-particle vertex functions. 
(Note that recently, PFFRG formulations with an initial guess of the self energy term $\Sigma^\infty \neq 0$ have also been 
developed~\cite{Reuther-2014}.) The PFFRG {\it Ansatz} (for recent reviews on FRG, see Refs.~\onlinecite{Metzner-2012,Platt-2013}) 
formulates an infinite hierarchy of coupled integrodifferential equations for the evolution of all $m$-particle vertex 
functions under the flow of $\Lambda$. Within PFFRG, the truncation of this system of equations to a closed set is accomplished 
by the inclusion of only two-particle reducible two-loop contributions, which are found to ensure sufficient back-feeding of the 
self-energy corrections to the two-particle vertex evolution~\cite{Katanin-2004}. A crucial advantage of the PFFRG as compared 
to Abrikosov-type spin random phase approximation methods is that the diagrammatic summation installed by the flow incorporates 
vertex corrections between all interaction channels, i.e., the two-particle vertex includes graphs that favor magnetic order and 
those that favor disorder in such a way that it treats both tendencies on equal footing. A numerical solution of the PFFRG 
equations is made possible by (i) discretizing the frequency dependencies of the vertex functions and (ii) limiting the spatial 
dependencies to a finite cluster. In our calculations, the number of discretized frequencies is $64$, and the spatial extent is 
restricted to $\sim10$ lattice spacings. When some correlations extend beyond this limiting distance, we observe a change in the 
nature of the renormalization flow from smooth to an unstable one accompanied by enhanced oscillations. We {\it interpret} this 
behavior as the existence of a nearby spontaneous symmetry breaking phase. Reversely, the existence of a stable solution indicates 
the absence of long-range order. From the effective low-energy two-particle vertex, we obtain the real-space zero-frequency spin 
susceptibility $\alpha_{ij}$:
\begin{equation}
\alpha_{ij}=\int_{0}^{\infty}d\tau \langle S_{i}^{z}(\tau)S_{j}^{z}(0) \rangle.
\label{eqn:spin-susc1}
\end{equation}
From it, we get the momentum resolved spin susceptibility $\alpha(\mathbf{q})$ by a Fourier transform:
\begin{equation}
\alpha(\mathbf{q})=\frac{1}{N} \sum_{i,j} e^{\imath \mathbf{q}\cdot({\bf R}_{i}-{\bf R}_{j})}\alpha_{ij},
\label{eqn:spin-susc2}
\end{equation}
where $N$ is the size of the cluster used in real space.

In the case of a magnetically disordered regime, we can track possible valence-bond crystal and nematic orders by computing their 
respective response functions. Here, we are particularly interested in studying the tendency of the spin liquid towards 
spontaneous breaking of rotation symmetry, i.e., a nematic spin liquid. Within our PFFRG framework, a conceptually simple way to 
calculate the nematic response function $\kappa_{\rm nem}$, which measures the tendency of the system to support nematic order, 
is to add a small perturbation to the bare Hamiltonian which enters the flow equations as the initial condition for the 
two-particle vertex:
\begin{equation}\label{eqn:nem}
{\cal H}_{\rm nem}=\delta\sum_{(i,j)\in S}\mathbf{S}_{i}\cdot \mathbf{S}_{j}- 
                   \delta\sum_{(i,j)\in W}\mathbf{S}_{i}\cdot \mathbf{S}_{j},
\end{equation}
which strengthens the couplings $J_{ij}$ on all bonds in $S$ [$J_{ij}\rightarrow J_{ij}+\delta$ for $(i,j)\in S$] and weakens 
the couplings in $W$ [$J_{ij}\rightarrow J_{ij}-\delta$ for $(i,j)\in W$]. The bond pattern $P\equiv\{S_p,W_p\}$ employed here 
specifies the channel of lattice point group symmetry breaking one intends to investigate. Such a modification amounts to 
changing the initial conditions of the RG flow at large cutoff scales $\Lambda$. As $\Lambda$ is lowered, we keep track of the 
evolution of all NN spin susceptibilities $\alpha_{ij}$. We then define the $p$-pattern nematic response function for a given pair 
of adjacent sites $(i,j)$ by 
\begin{equation}\label{eqn:nem-response}
\kappa_{\rm nem}^P=\frac{J}{\delta}\frac{\alpha_{S_P}-\alpha_{W_P}}{\alpha_{S_P}+\alpha_{W_P}},
\end{equation}
where the normalization factor $J/\delta$ ensures that the RG flow starts with an initialized value of $\kappa_{\rm nem}^{P}=1$. 
If the absolute value $\kappa_{\rm nem}^P$ remains small under the RG flow, the system tends to equalize, i.e., to reject the 
perturbation on that link, while, if $\kappa_{\rm nem}^P$ develops a large value under the RG flow it indicates that the system 
supports the $p$-pattern nematic order. 

\begin{figure}
\includegraphics[width=1.0\columnwidth]{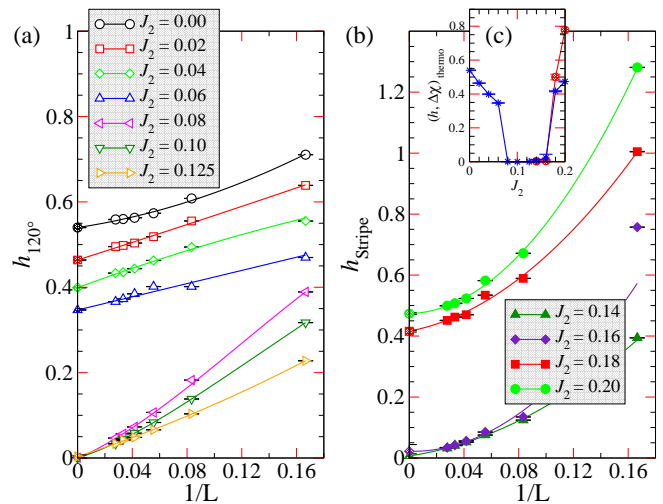}
\caption{\label{fig:fig4}
(Color online) Finite-size scalings of the antiferromagnetic variational parameter $h$ for (a) $120\degree$ coplanar magnetic order 
and (b) collinear stripe order are shown as a function of $1/L$, for different values of $J_2$. A quadratic fit is used in all cases 
and the cluster sizes correspond to $L=6$, $12$, $18$, $24$, $30$, and $36$. In (c), the thermodynamic estimate of $h$ (blue 
asterisks) and the nematic order parameter $\chi^{\rm NN}_{(1,1)}-\chi^{\rm NN}_{(1,0)}$ (red circles) are shown as functions of 
$J_2$. It is worth noting that nematic order onsets simultaneously with the stripe magnetic order. In panel (a) [A,P] boundary 
conditions are used for all clusters. In panel (b) [P,P] boundary conditions are used for $L=6$, $18$, and $30$ and [A,A] for $L=12$,
$24$, and $36$.}
\end{figure}

\section{Results}\label{sec:results}

\subsection{Variational energies}
Our variational calculations have been performed on $L \times L$ clusters possessing all symmetries of the triangular lattice, 
with periodic boundaries in the spin Hamiltonian of Eq.~(\ref{eqn:heis-ham}). Within the class of fully symmetric (preserving 
space group, time-reversal, and spin rotation symmetries) mean-field {\it Ans\"atze} with nonzero NN amplitudes, there exist only two 
$U(1)$ states. (i) The uniform resonating-valence bond (uRVB) state (for which all NN hoppings are real, equal in magnitude, and 
positive, leading to no magnetic fluxes piercing the lattice). This is a gapless state with a spinon Fermi surface. (ii) The DSL 
(for which all NN hoppings are real, of equal magnitude, and with the sign structure of Fig.~\ref{fig:fig3}(a), giving rise to 
$\pi$ flux in half of the triangles). This is a gapless state with low-energy Dirac conical excitations~\cite{Lu-2015}. The 
$L \times L$ clusters may be divided in two different classes, those with $L=4n$, or those with $L=4n+2$, where $n$ is a positive
integer. For the uRVB state, on both these cluster types, only [P,A] BC gives a closed shell configuration and is used for the 
present study. The other three BCs give an open shell. For the DSL, on $L=4n$ type clusters, [P,P] gives an open shell, the rest 
three give a closed shell and we use [A,P] BC in the present study. On $L=4n+2$ type clusters, all four boundary conditions give 
a closed shell, and we use [A,P] BC in the present study.

\begin{figure}
\includegraphics[width=1.0\columnwidth]{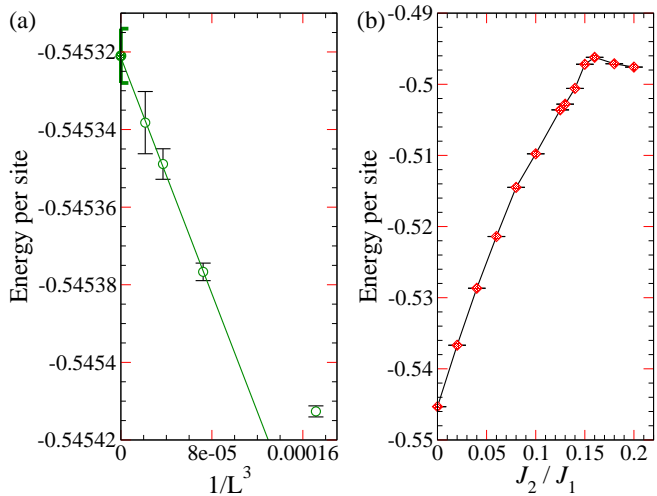}
\caption{\label{fig:fig5}
(Color online) (a) At $J_2=0$, the energies for $L=18$, $24$, $30$, and $36$ clusters are plotted. Assuming spin wave size 
scaling~\cite{Neuberger-1989,Azaria-1993}, we use the three largest clusters to get an estimate in the thermodynamic limit. 
Similarly, for other values of $J_2/J_1$, using the three largest clusters we perform size-scalings which are assumed to be linear 
in $1/{\rm L}^3$ for the antiferromagnetic regimes and $1/L$ for the spin-liquid regime, in order to obtain the corresponding 
thermodynamic estimates of the energy. They are plotted in (b) as a function of $J_2/J_1$. Note that here we use Jastrow wave function 
of Eq.~(\ref{eqn:physical-wf}) also for the spin-liquid regime.} 
\end{figure}

\floatsetup[table]{capposition=top}
\begin{table}
\begin{footnotesize}
\centering
\begin{tabular}{lll}
 \hline \hline
       \multicolumn{1}{c}{Method}
    & \multicolumn{1}{c}{$6\times6$ torus}
    & \multicolumn{1}{c}{$2D$ limit} \\ 
\hline \hline
       
\multirow{1}{*} VMC [see Eqs.~(\ref{eqn:physical-wf}) and~(\ref{eqn:mf-mag})]$^{*}$ & $-0.548025(3)$ & $-0.545321(7)$ \\
\multirow{1}{*} DMRG$^{*}$ & $-0.560375(6)$ & $-0.551(2)$  \\ 
\multirow{1}{*} Exact Diagonalization (Ref.~[\onlinecite{Bernu-1994}]) & $-0.5603734$ & $-0.5445$  \\ 
\multirow{1}{*} VMC (Heidarian {\it et al.}) (Ref.~[\onlinecite{Heidarian-2009}]) & $-0.55148(5)$ & $-0.5470(1)$ \\ 
\multirow{1}{*} VMC (Kaneko {\it et al.}) (Ref.~[\onlinecite{Kaneko-2014}]) & $-0.55519(4)$ & $-0.5449(2)$  \\ 
\multirow{1}{*} VMC (Weber {\it et al.}) (Ref.~[\onlinecite{Weber-2006}]) & $-0.543(1)$ & $-0.532(1)$  \\
\multirow{1}{*} FN (Yunoki {\it et al.}) (Ref.~[\onlinecite{Sorella-2006}]) & $ $ & $-0.53989(3)$ \\
\multirow{1}{*} FNE (Yunoki {\it et al.}) & $ $ & $-0.54187(6)$  \\ 
\multirow{1}{*} CCM (Li {\it et al.})$^{\dagger}$ (Ref.~[\onlinecite{Li-2015}]) & $ $ & $-0.5521(2)$  \\ 
\multirow{1}{*} SW (Chernyshev {\it et al.})$^{\dagger}$ (Ref.~[\onlinecite{Chernyshev-2009}]) & $ $ & $-0.54684$  \\
\multirow{1}{*} GFMC (Capriotti {\it et al.})$^{\dagger}$ (Ref.~[\onlinecite{Capriotti-1999}]) & $-0.5581(1)$ & $-0.5458(1)$  \\ 
\hline \hline

\end{tabular}
\end{footnotesize}
\caption{At $J_2=0$, estimations of the ground-state energy per site (in units of $J_1$) obtained by different methods on the
$6\times6$ torus and the thermodynamic limit. The energies computed by the first two methods (denoted by an asterisk) are from the 
present study. The DMRG estimate on the $6 \times 6$ torus is obtained by extrapolating in the truncation error, where at most 
$8000$ $SU(2)$ states are kept, while the thermodynamic estimate is obtained from extrapolating cylindrical clusters. After ED 
results, the next three rows show results from VMC studies. The first two employed a full pairing function optimization, the 
third one considered a wave function mixing $d_{x^2-y^2}+id_{xy}$ pairing and antiferromagnetic order. The results for fixed-node 
(FN), FN with an effective Hamiltonian (FNE), coupled cluster method (CCM), spin wave (SW), and Green's function Monte Carlo 
(GFMC) method with stochastic reconfiguration are also provided for comparison. Notice that the energies computed by the last three 
methods (denoted by a dagger) are not rigorous upper bounds of the ground-state energy.}
\label{tab:en-j2_00}
\end{table}

The variational energies of these two states provide a reference for all the rest of the paper and are given (on the 
$32 \times 32$ cluster) by (here the spin-spin Jastrow factor is not included)
\begin{eqnarray}\label{eqn:EnergyVsJ2a}
E_{\rm uRVB}/J_1 &=& -0.35446(1) + 0.01741(1)~J_2/J_1, \\
E_{\rm DSL}/J_1  &=& -0.52905(1) + 0.21657(1)~J_2/J_1.
\label{eqn:EnergyVsJ2b}
\end{eqnarray}
Hence the DSL energetically outshines the uRVB state for the parameter regime $0\leqslant J_2/J_1 \leqslant 0.8766(2)$. 

Let us now discuss our VMC results. First of all, we investigate the region with $0 \leqslant J_2/J_1 \leqslant 1/8$. Here, we 
consider the variational wave function that is generated from the uncorrelated Hamiltonian~(\ref{eqn:mf-mag}) with 
$\mathbf{q}=(4\pi/3,0)$ and the hopping {\it Ansatz} of Fig.~\ref{fig:fig3}(a). The variational parameters are the antiferromagnetic 
parameter $h$ and all the independent $u_{ij}$'s of the spin-spin Jastrow factor~(\ref{eqn:jastrow}). After optimizing on cluster 
sizes up to $L=36$, we find that for $J_2/J_1 \leqslant 0.06$, the $h$ parameter extrapolates to a finite value in the thermodynamic 
limit, indicating the existence of magnetic order, see Fig.~\ref{fig:fig4}(a). By contrast, for $J_2/J_1 \geqslant 0.08$, the strong 
frustration is able to melt the antiferromagnetic order; here, the $h$ parameter extrapolates to zero (within error-bars) in the 
thermodynamic limit, see Fig.~\ref{fig:fig4}(a). The high-accuracy of our magnetic wave functions is demonstrated by the fact that 
for $J_2=0$, i.e., the NN Heisenberg antiferromagnet, we get an energy per site in the thermodynamic limit of $E/J_1=-0.545321(7)$, 
which ranks amongst the most competitive variational estimates reported so far, see Fig.~\ref{fig:fig5}(a) and 
Table~\ref{tab:en-j2_00} for comparison of energies from different methods.

Then, we move towards investigating the second regime of interest, i.e., $1/8 \leqslant J_2/J_1 \leqslant 1$. Here, the ordering 
wave vector is expected to be that of stripe collinear order, i.e., $\mathbf{q}=(0,2\pi/\sqrt{3})$, which breaks rotation 
symmetry; hence, a consistent choice of the {\it Ansatz} for $\chi_{ij}$ in Eq.~(\ref{eqn:mf-mag}) is the one of Fig.~\ref{fig:fig3}(b).
It is found upon optimization on cluster sizes up to $L=36$, that for $J_2/J_1\leqslant 0.16$, the $h$ parameter extrapolates to 
zero (within error-bars) in the thermodynamic limit. Instead, for $J_2/J_1\geqslant 0.16$, the frustration starts getting 
relieved, and the collinear order onsets for $J_2/J_1\geqslant 0.18$, see Fig.~\ref{fig:fig4}(b). 

Therefore, our results strongly suggest that close to the classical transition $J_2/J_1=1/8$ a quantum paramagnet settles down for
$0.06 \lesssim J_2/J_1 \lesssim 0.16$. This estimation of the phase boundaries is in excellent agreement with previous 
DMRG~\cite{White-2015,Hu-2015} and CCM~\cite{Li-2015} studies. Moreover, by a close inspection of the thermodynamic energy per site 
versus $J_2/J_1$, we obtain a quite clear evidence that the first transition is continuous (no visible discontinuities on the energy),
while the second one is first-order, see Fig.~\ref{fig:fig5}(b). 

\floatsetup[table]{capposition=bottom}
\begin{table}[b]
\centering
\begin{tabular}{lcccc}
 \hline \hline
       \multicolumn{1}{l}{State}
    & \multicolumn{1}{c}{Gapped?}
    & \multicolumn{1}{c}{Unit cell}
    & \multicolumn{1}{c}{Parent state} 
    & \multicolumn{1}{c}{$\#$ in Ref.~[\onlinecite{Lu-2015}]} \\ 
\hline \hline

\multirow{1}{*} uRVB                       & No  & $1\times1$ &       &  \\ 
\multirow{1}{*} DSL                        & No  & $1\times2$ &       &  \\       
\multirow{1}{*} Z$_{2}\{0\}\mathcal{A}$    & Yes & $1\times1$ & uRVB  & $1$ \\
\multirow{1}{*} Z$_{2}\{\pi\}\mathcal{A}$  & Yes & $1\times2$ & DSL   & $20$ \\ 
\multirow{1}{*} Z$_{2}~\mathcal{C}$        & Yes & $1\times2$ & None  & $6$ \\ 
\multirow{1}{*} Z$_{2}\{\pi\}\mathcal{B}$  & No  & $1\times2$ & DSL   & $18$ \\  
\multirow{1}{*} VBC$_{2}$                  & Yes & $1\times2$ & DSL   &  \\  
\multirow{1}{*} VBC$_{4}$                  & Yes & $2\times2$ & DSL   &  \\  
\hline \hline

\end{tabular}
\caption{A list of states whose wave functions are studied.}
\label{tab:states}
\end{table}

\begin{figure*}
\includegraphics[width=1.0\columnwidth]{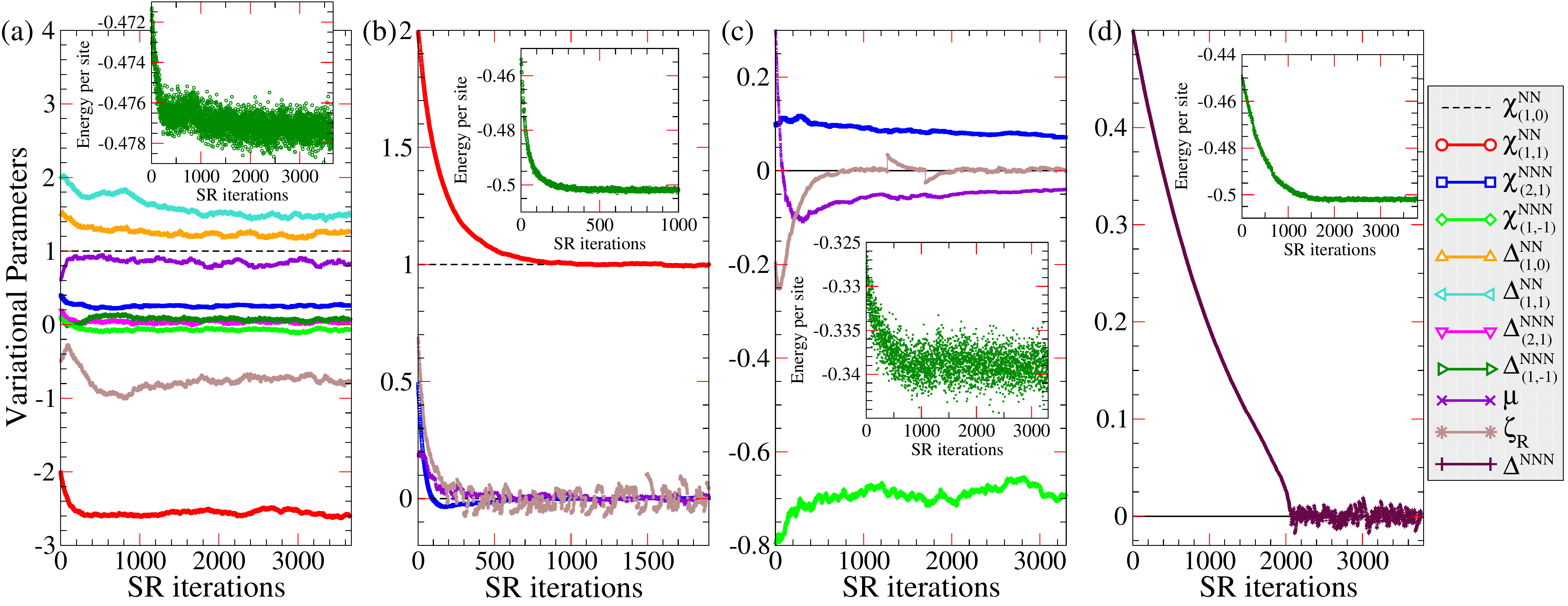}
\caption{\label{fig:fig6}
(Color online) A typical variational Monte Carlo optimization run for the promising candidate gapped nematic (a)${-}$(c), and gapless 
symmetric (d) $\mathbb{Z}_{2}$ spin-liquid wave functions, is shown for $J_2/J_1=1/8$ on the $L=32$ cluster. The variational 
parameters and energy per site (insets) are shown as a function of the Monte Carlo iterations. 
For (a) $\mathbb{Z}_{2}\{0\}\mathcal{A}$ spin liquid and (c) $\mathbb{Z}_{2}\mathcal{C}$ spin liquid, a gapped nematic 
$\mathbb{Z}_{2}$ state is stabilized albeit with a much higher energy compared to the DSL. For (b) $\mathbb{Z}_{2}\{\pi\}\mathcal{A}$
and (d) $\mathbb{Z}_{2}\{\pi\}\mathcal{B}$ states, the optimization yields the DSL. The optimized parameter and energy values are 
obtained by averaging over a much larger number of converged Monte Carlo iterations than shown above.}
\end{figure*}

We now address the important question of identifying the nature of the quantum paramagnet and study competing candidate states 
(see Table~\ref{tab:states}). In the following, we consider spin-liquid states generated from the uncorrelated 
Hamiltonian~(\ref{eqn:MF-SL}) {\it without} the spin-spin Jastrow factor. Indeed, although the inclusion of the latter one gives a 
slight improvement of the variational energy, it spoils the spin $SU(2)$ invariance of the correlated wave function. Instead, here 
we want to deal with {\it bona fide} spin-liquid wave functions. Following the previous work done by 
Zheng {\it et al.}~\cite{Zheng-2015} and Lu~\cite{Lu-2015}, we have analyzed all the symmetric (gapped and gapless) $\mathbb{Z}_{2}$ 
spin liquids and their nematic counterparts which can be realized with finite NN and NNN mean-field amplitudes. Among the ones that 
have been classified in Ref.~[\onlinecite{Lu-2015}], we list those that show some relevance after numerical optimization.

\begin{enumerate}
\item The $\mathbb{Z}_{2}\{0\}\mathcal{A}$ state: it has uniform hoppings and pairings at NN and NNN amplitudes, in addition, 
chemical potential and on-site pairing are also allowed. The uniform pairing terms can potentially open up a gap on the spinon 
Fermi surface formed by NN hoppings, leading to a symmetric gapped $\mathbb{Z}_{2}$ spin liquid, whose nematic version just 
differs in having two different classes of NN [$(1,0)$ and $(1,1)$] and NNN [$(2,1)$ and $(1,-1)$] hoppings and pairings, see 
Fig.~\ref{fig:fig3}(a). Hence, this {\it Ansatz} is continuously connected to the uRVB state, and is the only gapped one in its 
neighborhood. This state corresponds to the {\it Ansatz} No.~$1$ of Ref.~[\onlinecite{Lu-2015}].

\item The $\mathbb{Z}_{2}\{\pi\}\mathcal{A}$ state: the symmetric version has NN hoppings equal to the ones of the DSL, with the 
NNN hoppings being identically zero by symmetry, and an on-site pairing term that opens a gap at the Dirac nodes. The nematic 
version of the spin-liquid {\it Ansatz} is shown in Fig.~\ref{fig:fig3}(b). It is continuously connected to the DSL and happens to be 
the only gapped spin liquid in its neighborhood. This state corresponds to the {\it Ansatz} No.~$20$ in Ref.~[\onlinecite{Lu-2015}].

\item The $\mathbb{Z}_{2}\mathcal{C}$ state: the symmetric version has {\it all} identically vanishing NN amplitudes and is 
neither continuously connected to the uRVB nor to the DSL. The {\it Ansatz} for the nematic version of this spin liquid is shown in 
Fig.~\ref{fig:fig3}(c). This spin liquid corresponds to the {\it Ansatz} No.~$6$ of Ref.~[\onlinecite{Lu-2015}].

\item The $\mathbb{Z}_{2}\{\pi\}\mathcal{B}$ state: this is a gapless $\mathbb{Z}_{2}$ spin liquid which is continuously connected 
to the DSL. Its symmetric version is derived by adding a NNN pairing term on top of NN hopping $\pi$-flux {\it Ansatz}, which breaks the 
$U(1)$ gauge structure down to $\mathbb{Z}_{2}$ but keeps the spinon spectrum gapless. This state corresponds to the {\it Ansatz} No.~$18$ 
of Ref.~[\onlinecite{Lu-2015}].
\end{enumerate}

Let us focus our attention to the case with $J_2/J_1=1/8$ (well inside the magnetically disordered region); the energies per site 
of the two parent gapless $U(1)$ states are (on the $32 \times 32$ cluster) $E_{\rm uRVB}/J_1=-0.352287(3)$ and 
$E_{\rm DSL}/J_1=-0.501980(1)$, see Eqs.~(\ref{eqn:EnergyVsJ2a}) and~(\ref{eqn:EnergyVsJ2b}). The results of optimizations of the 
above four competing $\mathbb{Z}_{2}$ spin liquids wave functions on a $32 \times 32$ cluster are now discussed. For a generic 
unbiased starting point in the variational space of the $\mathbb{Z}_{2}\{0\}\mathcal{A}$ spin liquid, the variation of parameters 
and energy in the SR optimization is shown in Fig.~\ref{fig:fig6}(a). It is found that the parent uRVB spin liquid undergoes a 
pairing instability towards this gapped nematic $\mathbb{Z}_{2}$ state, which is stabilized with an optimal energy of 
$E/J_1=-0.47729(2)$. The energy gain is significant, but not enough to overcome the DSL. This situation should be contrasted with 
the spin-$1/2$ kagome Heisenberg antiferromagnet, where the spinon Fermi surface state remains stable towards all gapped 
$\mathbb{Z}_{2}$ spin liquids in its neighborhood~\cite{Iqbal-2011b,Iqbal-2015a}. 

We now discuss the fate of the $\mathbb{Z}_{2}\{\pi\}\mathcal{A}$ spin liquid, whose proximity to the DSL makes it the most 
promising candidate state on energetic grounds. The parameter and energy optimization in Fig.~\ref{fig:fig6}(b) clearly show that 
the energy converges neatly to the reference value of the DSL corresponding to $\chi^{\rm NN}_{(1,1)} \rightarrow 1$ and 
($\chi^{\rm NNN}_{(2,1)}$, $\mu$, $\zeta_{\rm R}$)$\rightarrow 0$. Here, we would like to bring attention to the important fact 
that despite the energy having essentially converged after $\approx400$ iterations, the parameters did not converge and were 
still varying, converging to their final values much later than the energy. This fact is possible because, in the energy 
minimization, forces are calculated through the correlated sampling and not by energy differences~\cite{Sorella-2005}. The energy 
landscape along the manifold connecting the DSL to the $\mathbb{Z}_{2}\{\pi\}\mathcal{A}$ one is thus very flat. Consequently, 
assessing the stability of the DSL by solely computing the energy of the perturbed wave function with fixed parameters (i.e., point 
by point independently, as was done in Ref.~[\onlinecite{Zheng-2015}]) is very hard. Only by performing an accurate SR optimization 
method can one successfully optimize the parameters and transparently show that $\chi^{\rm NN}_{(1,1)}=1$ and 
($\chi^{\rm NNN}_{(2,1)}$, $\mu$, $\zeta_{\rm R}$)$=0$ corresponds to the actual minimum of the variational energy. This fact 
implies that the nematic order parameter is absent, the system is gapless, with the $U(1)$ gauge structure kept intact, and hence,
the DSL is {\it locally} and {\it globally} stable with respect to destabilizing into the $\mathbb{Z}_{2}\{\pi\}\mathcal{A}$ state. 
We verified this result by doing many optimization runs starting from different random initial values of the parameters in the 
four-dimensional variational space. This situation is in consonance with what happens for the kagome Heisenberg 
antiferromagnet~\cite{Iqbal-2011b}. 

The $\mathbb{Z}_{2}\mathcal{C}$ state suffers the same fate as the $\mathbb{Z}_{2}\{0\}\mathcal{A}$ spin liquid, namely that upon 
optimization [see Fig.~\ref{fig:fig6}(c)], it is stabilized but with a much higher energy of $E/J_1=-0.33877(3)$ compared to the 
DSL. Lastly, we study the gapless symmetric $\mathbb{Z}_{2}\{\pi\}\mathcal{B}$ state. Also in this case, we find that the 
optimization smoothly goes back to the DSL with $\Delta^{\rm NNN} \rightarrow 0$, see Fig.~\ref{fig:fig6}(d). For all the above 
four spin liquid wave functions, we reach similar conclusions for various values of $J_2/J_1$ (within the spin liquid regime), 
other cluster sizes, and different choices of allowed boundary conditions.

\begin{figure}[t]
\includegraphics[width=1.0\columnwidth]{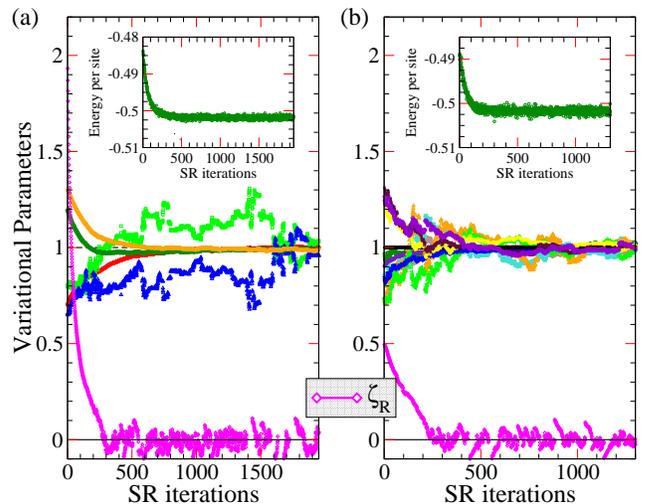}
\caption{\label{fig:fig7}
(Color online) A typical variational Monte Carlo optimization is shown, at $J_2/J_1=1/8$ on a $32\times32$ cluster, for (a) 
two-site unit cell VBC (six independent bonds) and (b) four-site unit cell VBC (12 independent bonds) wave functions. The variational 
parameters and energy per site (insets) are shown as a function of the Monte Carlo iterations. On starting from different sets 
of initialized parameter values, for both cases the optimization gives back the DSL, with all hoppings being equal and 
$\zeta_{\rm R}=0$ ($\chi_{1}=1$ is set as the reference). Note that we include the on-site pairing $\zeta_{\rm R}$ so as to capture 
potential instabilities of the $\mathbb{Z}_{2}\{\pi\}\mathcal{A}$ state towards VBC ordering.}
\end{figure}

\begin{figure*}
\includegraphics[width=1.0\columnwidth]{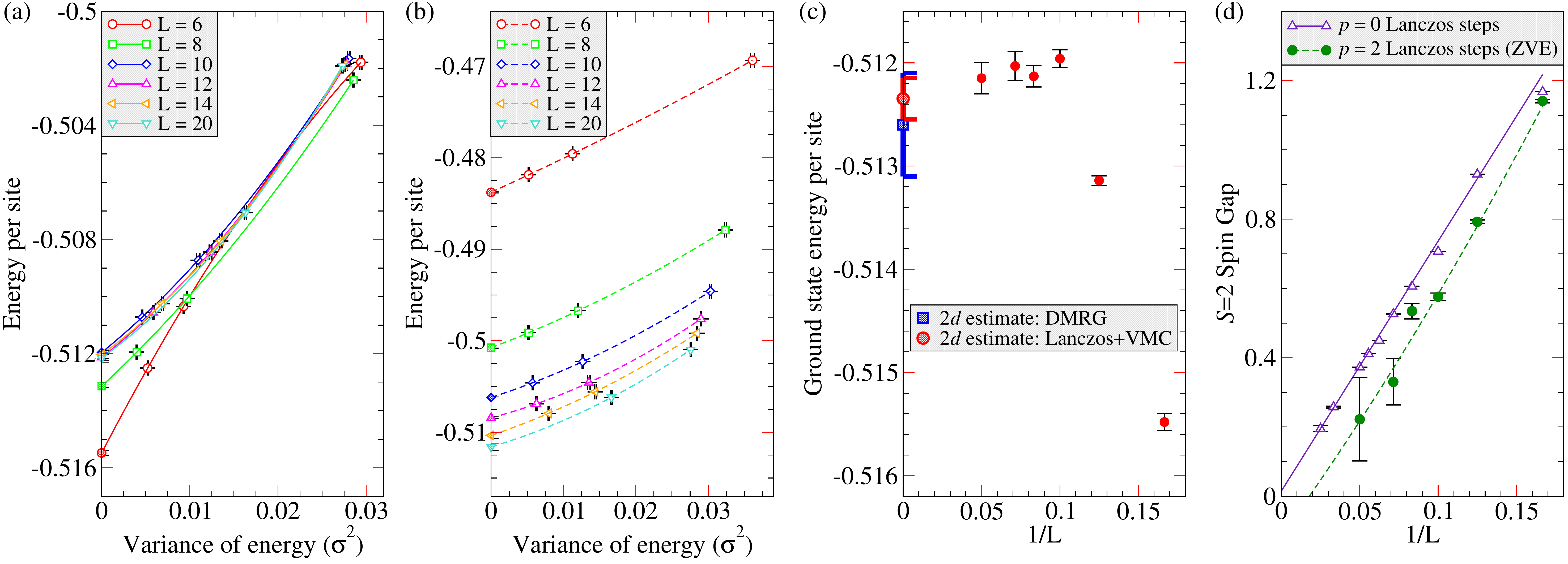}
\caption{\label{fig:fig8}
(Color online) At $J_2/J_1=1/8$ for the DSL, (a) Variational energies as a function of the variance of energy, for zero, one, and 
two Lanczos steps. The $S=0$ ground-state energy on the $L=6$, $8$, $10$, $12$, and $14$ clusters is estimated by extrapolating the 
three variational results to the zero-variance limit by a quadratic fit, while only two points have been used for the $L=20$ cluster 
(see the main text for a description of the extrapolation method). (b) The same for the $S=2$ excited state. (c) The thermodynamic 
estimate of the ground-state energy obtained by a finite-size scaling of the estimated ground-state energies. (d) Finite-size scaling
of the $S=2$ spin gap as a function of $1/L$. Both the $p=0$ and $p=2$ extrapolated values give an estimate which is zero (within 
error-bars) in the thermodynamic limit. A linear fit is used for $p=0$ and a quadratic fit for $p=2$. The largest cluster considered 
for $p=0$ has $L=40$.}
\end{figure*}

\floatsetup[table]{capposition=bottom}
\begin{table*}
\begin{footnotesize}
\centering
\begin{tabular}{llllllllll}
 \hline \hline
       \multicolumn{1}{c}{Size}
    & \multicolumn{1}{c}{$0$-LS}
    & \multicolumn{1}{c}{$1$-LS} 
    & \multicolumn{1}{c}{$2$-LS}
    & \multicolumn{1}{c}{$0$-LS}
    & \multicolumn{1}{c}{$1$-LS}
    & \multicolumn{1}{c}{$2$-LS}
    & \multicolumn{1}{c}{Ground state}  
    & \multicolumn{1}{c}{$S=2$ state} 
    & \multicolumn{1}{c}{$S=2$ gap}  \\ \hline
       
\multirow{1}{*}{$6$} & $-0.501788(1)$ & $-0.510344(2)$ & $-0.512503(3)$ & $-0.469360(2)$ & $-0.479580(1)$ & $-0.481870(3)$ & $\bm{-0.51548(8)}$ & $\bm{-0.48378(6)}$ & $\bm{1.141(5)}$  \\ 
                                                                                                                       
\multirow{1}{*}{$8$} & $-0.502410(1)$ & $-0.510070(1)$ & $-0.511950(4)$ & $-0.487890(1)$ & $-0.496720(1)$ & $-0.499120(5)$ & $\bm{-0.51314(4)}$ & $\bm{-0.50076(7)}$ & $\bm{0.792(5)}$  \\ 
                                                                                                                                          
\multirow{1}{*}{$10$} & $-0.501655(1)$ & $-0.508724(2)$ & $-0.510721(5)$ & $-0.494589(1)$ & $-0.502285(2)$ & $-0.50459(1)$ & $\bm{-0.51195(8)}$ & $\bm{-0.50619(6)}$ & $\bm{0.576(10)}$   \\

\multirow{1}{*}{$12$} & $-0.501828(1)$ & $-0.508438(3)$ & $-0.510558(5)$ & $-0.497619(1)$ & $-0.504574(2)$ & $-0.506895(4)$ & $\bm{-0.51213(10)}$ & $\bm{-0.50841(11)}$ & $\bm{0.53(2)}$  \\ 

\multirow{1}{*}{$14$} & $-0.501866(1)$ & $-0.508049(3)$ & $-0.510245(7)$ & $-0.499183(1)$ & $-0.505574(3)$ & $-0.507928(3)$ & $\bm{-0.51203(14)}$ & $\bm{-0.51034(30)}$ & $\bm{0.33(6)}$  \\

\multirow{1}{*}{$20$} & $-0.501915(1)$ & $-0.507056(3)$ & $$ & $-0.500984(1)$ & $-0.506190(3)$ & $$ & $\bm{-0.51214(15)}$ & $\bm{-0.5116(4)}$ & $\bm{0.22(12)}$    \\ \hline \hline

\end{tabular}
\end{footnotesize}
\caption{At $J_2/J_1=1/8$, the energies of the DSL (columns $2{-}4$) and its $S=2$ excited state (columns $5{-}7$), with $p=0$, 
$1$, and $2$ Lanczos steps on different cluster sizes obtained by VMC are given. The ground-state and $S=2$ excited-state energies
of the spin-$1/2$ Heisenberg model estimated by using zero-variance extrapolation of variational energies on different cluster 
sizes are marked in bold. The $S=2$ gap obtained from the estimates of $S=0$ and $S=2$ energies on different cluster sizes is 
given in the last column.}
\label{tab:en-lanczos}
\end{table*}

Valence-bond crystal (VBC) states also represent an important class of instabilities of spin liquids~\cite{Read-1989,Read-1990}. 
Here, we study VBC orders~\cite{Slagle-2014}, with two- and four-site unit cells, respectively called VBC$_{2}$ and VBC$_{4}$ 
states, and investigate if the DSL and $\mathbb{Z}_{2}\{\pi\}\mathcal{A}$ states are stable towards undergoing a phase transition 
to these VBC orders under perturbations. The result of optimizations of VBC wave functions are shown in Fig.~\ref{fig:fig7}. 
We find that both these spin-liquid {\it Ans\"atze} are locally and globally stable towards dimerizing into VBC orders, similar to the 
situation on the kagome lattice~\cite{Iqbal-2011a,Iqbal-2012}.

\begin{figure}
\includegraphics[width=1.0\columnwidth]{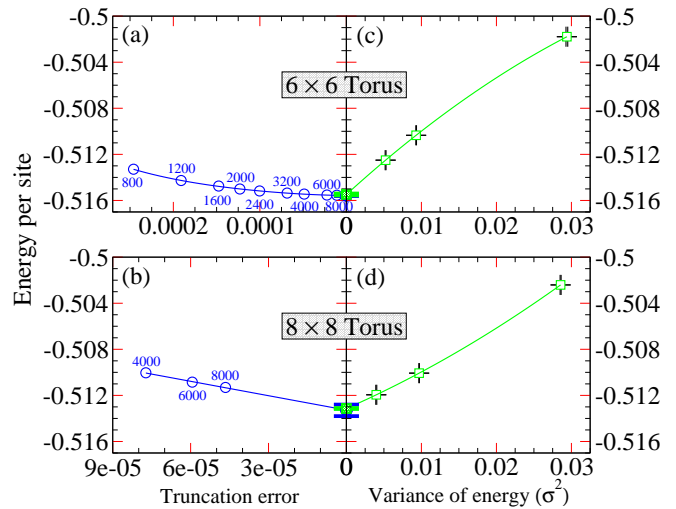}
\caption{\label{fig:fig9}
(Color online) At $J_2/J_1=1/8$, the estimate of the ground-state energy per site obtained by [(a) and (b)] DMRG and [(c) and (d)] VMC, on 
$6 \times 6$ and $8 \times 8$ tori [see Table~\ref{tab:en-j2_125}]. The values next to the DMRG data points correspond to the 
number of $SU(2)$ states kept.}
\end{figure}

Having established that the DSL is stable and represents the lowest energy variational wave function (within the class of 
Gutzwiller-projected fermionic states), we consider it as the starting ($p=0$) state and improve it by applying Lanczos 
steps~\cite{Iqbal-2013,Iqbal-2014,Iqbal-2015b}. At $J_2/J_1=1/8$, the effect of two Lanczos steps for different cluster sizes is 
shown in Fig.~\ref{fig:fig8}(a) [see also Table~\ref{tab:en-lanczos} for the actual values of the energies of the DSL]. Our estimate 
of the ground-state energy on $6 \times 6$ cluster is in excellent agreement with the exact result, and equal (within error-bars) to 
the corresponding estimate from DMRG [see Table~\ref{tab:en-j2_125}, and Figs.~\ref{fig:fig9}(a) and~\ref{fig:fig9}(c)]. Similarly, 
on the $8 \times 8$ cluster, the estimate of the ground-state energy from VMC is equal (within error-bars) to the DMRG estimate 
[see Table~\ref{tab:en-j2_125} and Fig.~\ref{fig:fig9}(b) and Fig.~\ref{fig:fig9}(d)]. Moreover, our VMC energies are considerably 
lower compared to the ones from  previous VMC studies [see Table~\ref{tab:en-j2_125}]. 

The very computationally demanding $p=2$ calculations have been performed for $L=6$, $8$, $10$, $12$, and $14$ clusters, while for 
the $L=20$ cluster only the first Lanczos step has been considered. In the former cases, the zero-variance extrapolation can be 
exploited by a quadratic fit of the three points, $E=E_{\rm ex}+\mathcal{A}\times\sigma^{2}+\mathcal{B}\times(\sigma^{2})^{2}$.
For the largest cluster, i.e., $L=20$, we also considered a quadratic fit: we first obtained an estimate of the $\mathcal{A}$ and 
$\mathcal{B}$ coefficients by a size scaling of the smaller clusters and then verified that indeed, these values give an excellent 
fit (i.e., least mean-square error) of the points for $L=20$ cluster. It is worth mentioning, that the zero-variance extrapolation 
gives size consistent results for the energy per site. Indeed, even though the Lanczos step procedure (with a fixed $p$) becomes 
less and less efficient when increasing the system size, the extrapolation procedure remains accurate: this can be seen by noticing
that the gain in the energy and variance with respect to $p=0$ decreases with $L$, but the extrapolation is not affected, since 
the slope is essentially unchanged~\cite{Hu-2013}.

By using the ground-state energy estimates on different cluster sizes, we performed a finite-size extrapolation, see 
Fig.~\ref{fig:fig8}(c). At $J_{2}/J_{1}=1/8$, our final estimate for the energy of the infinite two-dimensional system is:
\begin{equation}
E^{2D}_{\infty}/J_1=-0.51235(20).
\end{equation} 
This estimate is equal (within error-bars) with the DMRG estimate and much lower compared to VMC estimates in previous studies 
[see Table~\ref{tab:en-j2_125}]. It is worth mentioning that our energies are obtained with a state that has all the symmetries of
the lattice, while DMRG states are obtained on cylinders with open boundaries.

\floatsetup[table]{capposition=top}
\begin{table}[t]
\centering
\begin{tabular}{llll}
 \hline \hline
       \multicolumn{1}{l}{Method}
    & \multicolumn{1}{c}{$6\times6$ torus}
    & \multicolumn{1}{c}{$8\times8$ torus}
    & \multicolumn{1}{c}{$\infty~2D$ limit} \\ \hline
       
\multirow{1}{*} VMC$^{*}$ & $-0.51548(8)$ & $-0.51314(4)$ & $-0.51235(20)$ \\
\multirow{1}{*} DMRG$^{*}$ & $-0.51557(5)$ & $-0.5133(5)$ & $-0.5126(5)$  \\ 
\multirow{1}{*} ED$^{*}$ & $-0.515564$ & $$ & $$  \\ 
\multirow{1}{*} VMC (Ref.~[\onlinecite{Kaneko-2014}]) & $-0.5089(1)$ & $ $ & $-0.5028(2)$  \\  \hline \hline

\end{tabular}
\caption{At $J_2/J_1=1/8$, the estimates of the ground-state energy per site (in units of $J_{1}$) obtained by different methods 
on $6 \times 6$ and $8 \times 8$ tori, as well as its estimation in the thermodynamic limit. The VMC estimates are obtained using 
zero-variance extrapolation and the DMRG estimates by an extrapolation in truncation error (keeping at most $8000~SU(2)$ states), 
see Fig.~\ref{fig:fig9}. The thermodynamic estimate from DMRG is obtained by using cylindrical clusters~\cite{Hu-2015}. The asterisk 
denotes results from present study. For comparison, in the last row we show VMC results of Kaneko {\it et al.} 
(Ref.~[\onlinecite{Kaneko-2014}]) employing a full pairing function optimization.}
\label{tab:en-j2_125}
\end{table}

\subsection{Spin excitation gap}
We now address the important issue of the spin gap. An estimation of the spin gap can be performed by constructing excited states 
of the DSL. Here, we consider a state with $S=2$, which is particularly simple to handle, since it corresponds to a single 
determinant constructed by changing boundary conditions with respect to the $S=0$ state, in order to have four spinons in an 
eightfold degenerate single-particle shell at the chemical potential; then, a unique mean-field state is obtained by taking all 
these spinons with the same spin (so that the total wave function has $S=2$). The effect of applying two Lanczos steps on the 
$S=2$ excited state for different cluster sizes is shown in Fig.~\ref{fig:fig8}(b) [see Table~\ref{tab:en-lanczos} for the actual 
values of the energies]. 

Before Gutzwiller projection, the mean-field state is gapless by construction (there are Dirac cones in the mean-field spectrum); 
however, given the $U(1)$ low-energy gauge structure of the mean-field {\it Ansatz}, it is not obvious that this property is preserved 
after projection~\cite{Wen-2002}. In fact, the $U(1)$ gauge fluctuations are expected to be wild, possibly leading to some 
instability of the mean-field {\it Ansatz}~\cite{Hermele-2004}. Here, we show that the $S=2$ gap is vanishing in the thermodymanic limit,
even after including the Gutzwiller projection [see Fig.~\ref{fig:fig8}(d)], akin to the situation in the kagome Heisenberg 
model~\cite{Iqbal-2014,Iqbal-2015b}. The computation of the $S=0$ and $S=2$ energies allows us to obtain the {\it extrapolated} gap 
for each size independently, which is reported in Table~\ref{tab:en-lanczos} and Fig.~\ref{fig:fig8}(d). Here, despite having an 
error-bar that increases with the system size, we can reach an extremely accurate thermodynamic extrapolation, namely, the $S=2$ 
gap is $\Delta=-0.173 \pm 0.213$. Therefore, our main conclusion is that our competitive spin-liquid state has gapless excitations. 
Our best estimate for the upper bound on the $S=2$ gap is $\Delta \simeq 0.04$, leading to a $S=1$ gap that would be approximately 
half of this value, i.e., $\Delta_{\rm T} \simeq 0.02$. This latter result is considerably smaller than the DMRG estimate of 
$\Delta_{\rm T}=0.30(1)$~\cite{White-2015}, which was obtained by considering cylindrical geometries $L_x\times L_y$, i.e., 
first performing the limit $L_x\rightarrow\infty$ and then increasing the circumference $L_y$.

\begin{figure}
\includegraphics[width=1.0\columnwidth]{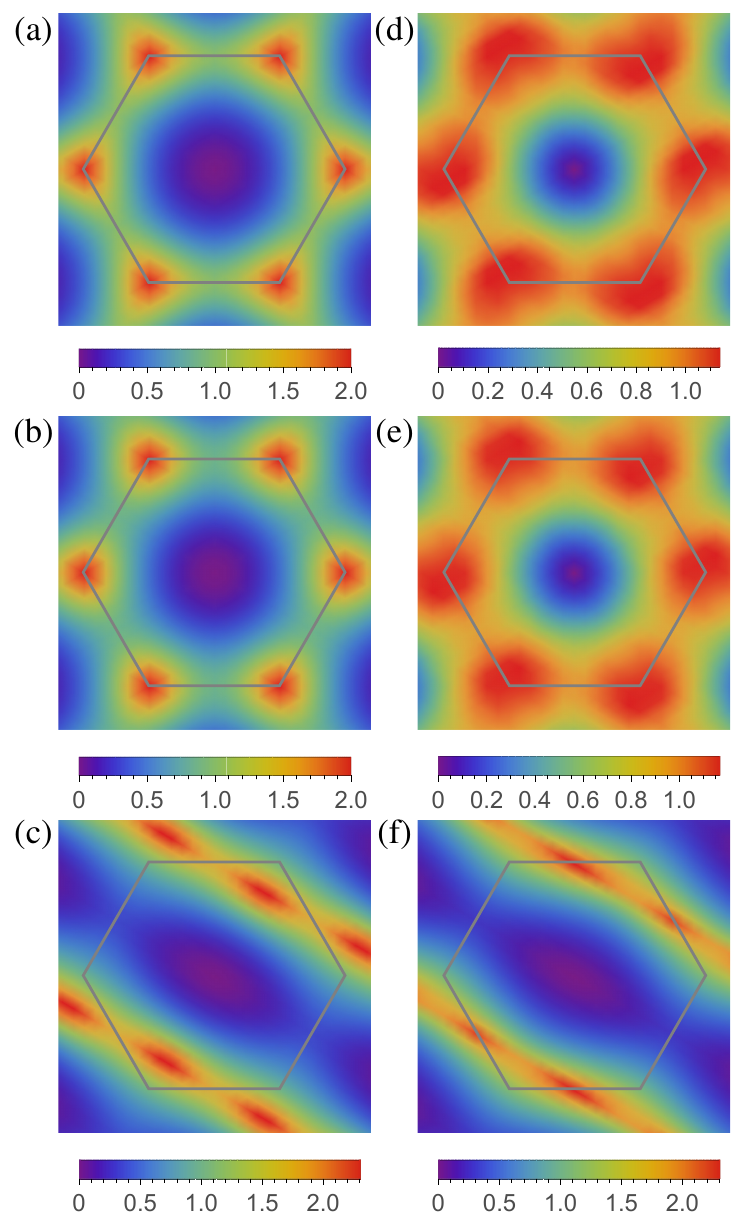}
\caption{\label{fig:fig10}
(Color online) Spin structure factors [Eq.~(\ref{eqn:sf})] obtained in VMC at $J_2/J_1=1/8$, for (a) DSL, (b) symmetric 
$\mathbb{Z}_{2}\{\pi\}\mathcal{A}$ state with imposed on-site pairing $\zeta_{\rm R}=2$ and (c) nematic 
$\mathbb{Z}_{2}\{\pi\}\mathcal{A}$ state with imposed nematicity $\chi_{(1,1)}/\chi_{(1,0)}=2.25$, which is approximately the value 
found in DMRG~\cite{Hu-2015}. Similarly, (d) uRVB spin liquid, (e) symmetric $\mathbb{Z}_{2}\{0\}\mathcal{A}$ state, and (f) nematic 
$\mathbb{Z}_{2}\{0\}\mathcal{A}$ state. For (e) and (f), the optimal parameters are determined by energy minimization.} 
\end{figure}

\subsection{Spin structure factor}
We finally consider the equal-time spin structure factor:
\begin{equation}\label{eqn:sf}
S(\mathbf{q}) = \frac{1}{N} \sum_{i,j} e^{-\imath \mathbf{q}\cdot ({\bf R}_i-{\bf R}_j)} 
\langle\mathbf{S}_{i} \cdot \mathbf{S}_{j}\rangle
\end{equation} 
where $N$ is the total number of sites. In Fig.~\ref{fig:fig10}, we present the results for $S(\mathbf{q})$ obtained within VMC for 
different competing variational wave functions. The gapless DSL shows well defined peaks at the corners of the Brillouin zone, 
which are, however, not related to the presence of a finite magnetic moment [i.e., $S(\mathbf{q})/N \to 0$ with $N \to \infty$]. The 
gapped symmetric $\mathbb{Z}_{2}\{\pi\}\mathcal{A}$ state displays a similar structure factor, even though the peaks are slightly 
broader and do not diverge in the thermodynamic limit. Finally, imposing nematic terms in the variational wave function, e.g., 
different hoppings along $(1,1)$ and $(1,0)$, easily introduces strong anisotropic features. A similar behavior is also observed 
in the uRVB state and its gapped and nematic descendants (see Fig.~\ref{fig:fig10}) albeit it is seen that the peaks are much 
more broader and diffused. From a qualitative point of view, it is seen that the spin structure profiles at large $\mathbf{q}$ for 
both the DSL and uRVB states bear a great deal of similarity to the profiles of their gapped symmetric descendants, which is a 
manifestation of the similar nature of short-distance correlations in these two distinct states. Nevertheless, while for gapped
states $S(\mathbf{q})$ remains finite in the thermodynamic limit, for gapless spin liquids $S(\mathbf{q})$ diverges for 
$\mathbf{q}=(4\pi/3,0)$ and symmetry-related points.

\begin{figure}
\includegraphics[width=1.0\columnwidth]{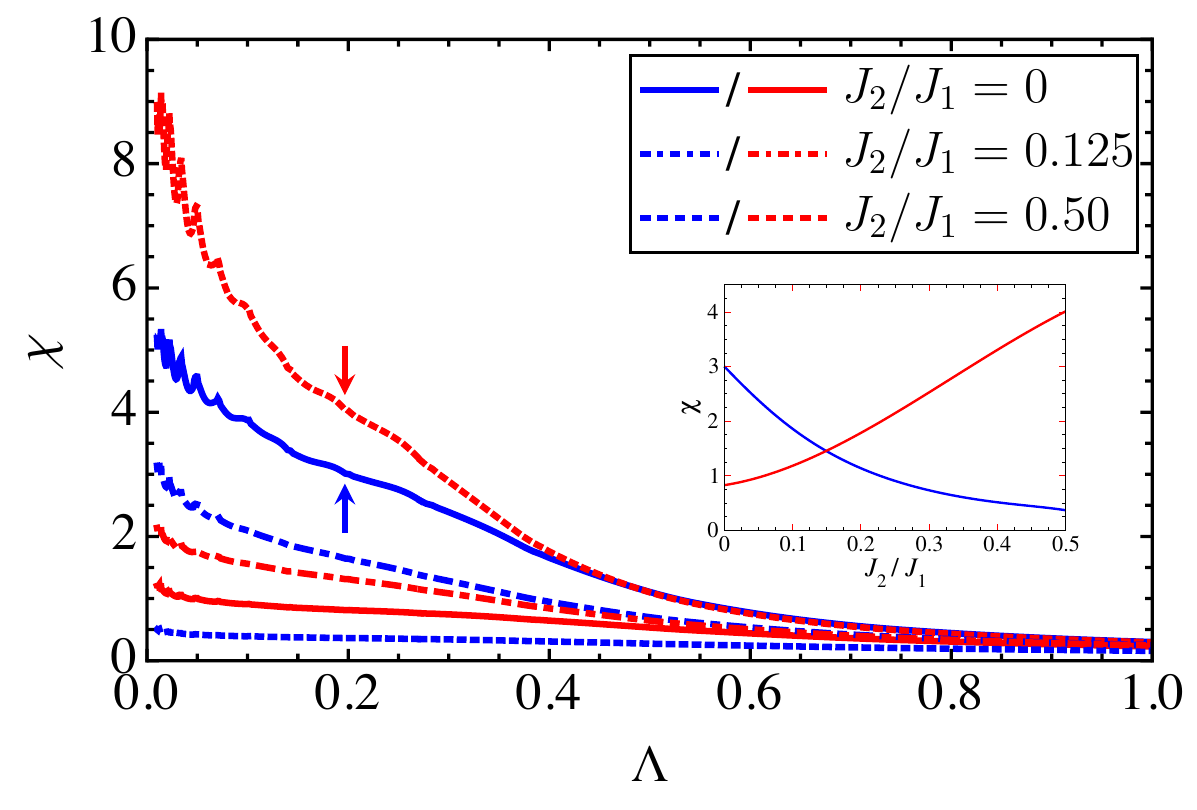}
\caption{\label{fig:fig11}
(Color online) RG flows of the momentum resolved spin susceptibility evaluated at the coplanar (blue) and collinear (red) ordering 
wave vectors for the three regimes of the phase diagram (Fig.~\ref{fig:fig1}). The arrows mark the beginning of the unphysical part 
of the flow induced by the onset of strong (finite-range) coplanar or collinear antiferromagnetic correlations. In the spin liquid 
regime the flow remains physical down to $\Lambda=0$. The small oscillations below $\Lambda \approx 0.1$ in this flow are due to 
frequency discretization. (Inset) For $\Lambda=0.2$, the susceptibility evaluated at the coplanar (blue) and collinear (red) 
ordering wave-vectors, as a function of $J_2/J_1$.} 
\end{figure}

\begin{figure*}
\includegraphics[width=0.995\columnwidth]{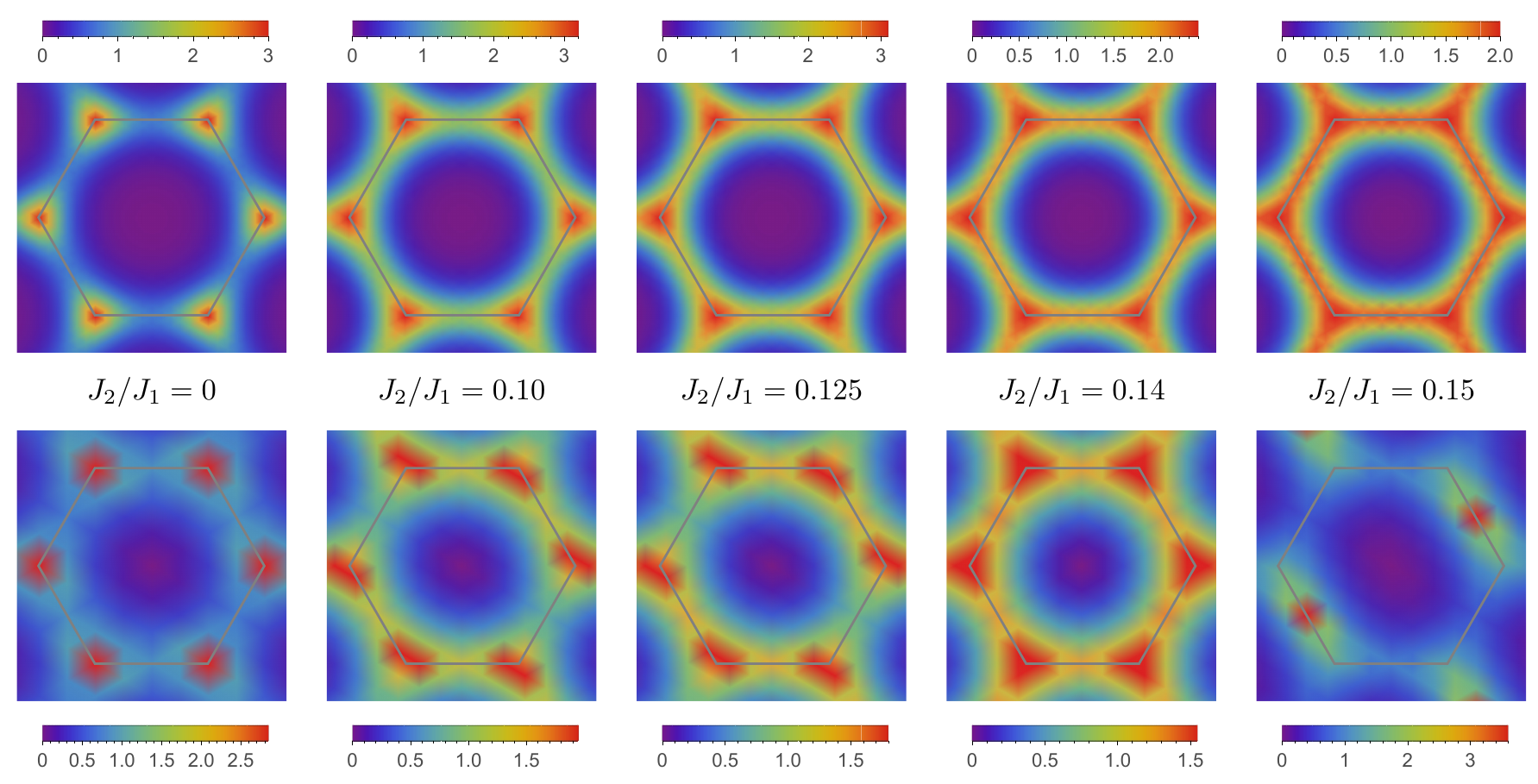}
\caption{\label{fig:fig12}
(Color online) (Top row) Momentum-resolved spin susceptibility [see Eq.~(\ref{eqn:spin-susc2})] profiles obtained in PFFRG. For 
$J_2/J_1=0$, evaluated at $\Lambda=0.20$ (marked by arrow in Fig.~\ref{fig:fig11}), and for other values of $J_2/J_1$, at 
$\Lambda=0$. (Bottom row) Spin structure factors [see Eq.~(\ref{eqn:sf})], as obtained by DMRG on cylindrical clusters with $L_y=6$ for 
$J_2=0$ and $L_y=8$ for the other values of $J_2/J_1$~\cite{Hu-2015}.} 
\end{figure*}

\begin{figure}[b]
\includegraphics[width=1.0\columnwidth]{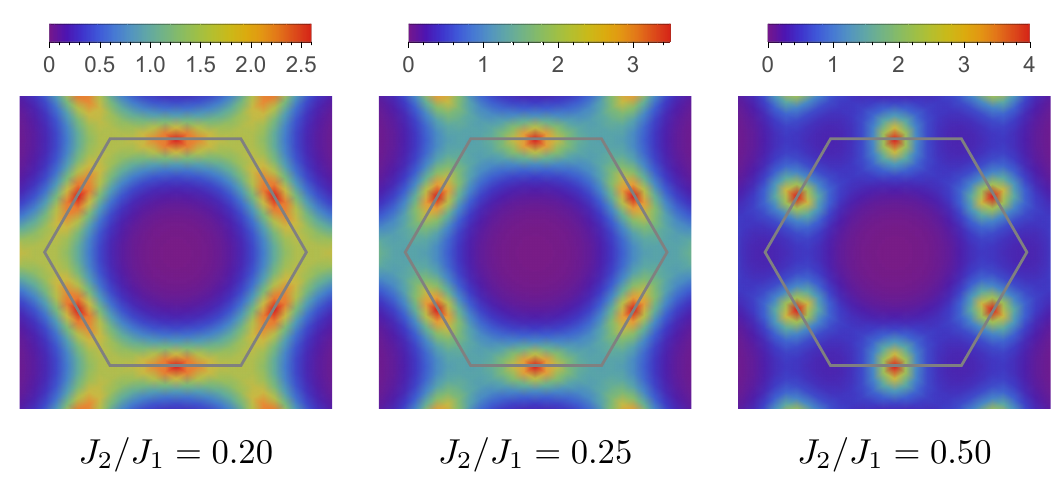}
\caption{\label{fig:fig13}
(Color online) Momentum-resolved spin susceptibility [see Eq.~(\ref{eqn:spin-susc2})] profiles obtained in PFFRG at larger values of 
$J_2/J_1$ for which strong collinear antiferromagnetic correlations set in (at finite distances). For $J_2/J_1=0.5$, evaluated at $\Lambda=0.20$ (marked by arrow in Fig.~\ref{fig:fig11}), and for $J_2/J_1=0.2$ and $0.25$, at $\Lambda=0$.} 
\end{figure}

A wide magnetically disordered phase is also obtained within PFFRG approach. As discussed in Sec.~\ref{sec:methods}, the 
presence or absence of magnetic order can be assessed from the behavior of the RG flow: in two-dimensional systems, the existence 
of long-range order at zero temperature is signaled by a breakdown of the RG flow, from a smooth to a rather unstable behavior; 
by contrast, a magnetically disordered phase at zero temperature is associated with a smooth RG flow down to $\Lambda=0$ (the 
presence of small oscillations at small values of $\Lambda$ are only due to the discretization of the frequencies). 
In Fig.~\ref{fig:fig11}, we report few relevant cases for the RG flow for the spin susceptibility of Eq.~(\ref{eqn:spin-susc2}). 
For small values of $J_2/J_1$ the largest susceptibility is the one with $\mathbf{q}=(4\pi/3,0)$, corresponding to a coplanar 
order; by contrast, for larger values of the frustrating ratio, the largest signal comes from $\mathbf{q}=(0,2\pi/\sqrt{3})$ that 
corresponds to stripe order. Most importantly, we find that for $0.04 \lesssim J_2/J_1 \lesssim 0.5$ the RG flow does not show any
tendency to instabilities down to $\Lambda=0$, indicating the presence of a quantum paramagnetic state at zero temperature, 
see Fig.~\ref{fig:fig11}. This result is also corroborated by the calculation of the full spin susceptibility in the whole 
Brillouin zone (this quantity is computed at $\Lambda=0$ when possible or just before the possible breakdown when some momentum 
shows an instability). The final results are shown in Fig.~\ref{fig:fig12} for $J_2/J_1 \leqslant 0.15$ and Fig.~\ref{fig:fig13} 
for larger values of $J_2/J_1$. For the former case, we also report, for comparison, the DMRG results for the equal-time structure
factor of Eq.~(\ref{eqn:sf}). Here, the computation is performed in long cylinders (i.e., $L_x \gg L_y$, with $L_y=6$ and $8$) 
where the spin-spin correlations are taken only inside the bulk (i.e., $6 \times 6$ and $8 \times 8$).  For small values of 
$J_2/J_1$, the spectral weight of both PFFRG and DMRG is concentrated at the corners of the Brillouin zone corresponding to 
coplanar $120\degree$ order. By increasing $J_2/J_1$ and entering the paramagnetic regime, a smearing of the weight is observed, 
with progressively diffused peaks as one goes deeper into the disordered regime. One important difference is that PFFRG gives a 
relatively broad paramagnetic regime, which extends from $J_2/J_1 \simeq 0.04$ to $J_2/J_1 \simeq 0.5$. The fact that PFFRG, due 
to frequency discretization and the current way of distinguishing paramagnetism from weak magnetic order, estimates a larger 
paramagnetic regime than VMC is consistent with previous results on the $J_1{-}J_2$ Heisenberg model on the kagome 
lattice~\cite{Suttner-2014}. By further increasing the frustrating ratio, the collinear antiferromagnetic phase is reached 
and the weight shifts along the boundary of the Brillouin zone, see Fig.~\ref{fig:fig13}. In DMRG, one observes that this shift of 
weight is very sudden upon going from $J_2/J_1=0.14$ to $J_2/J_1=0.15$. In contrast, in PFFRG the transfer of weight occurs much 
more gradually and starts for larger values of $J_2/J_1\sim 0.2$. Despite these quantitative deviations, the PFFRG is accurate 
in tracking the central correlation profiles, enabling a comparison across methods. In particular, PFFRG is seen to provide a 
reliable short-range correlation profile of a phase, whereas the estimation of long-distance correlations shows more quantitative
deviations.

To address the issue of the possible nematic nature of the quantum paramagnetic phase, we have computed within PFFRG the nematic 
response function $\kappa_{\rm nem}^{P}$ Eq.~(\ref{eqn:nem-response}). We observe that for any given ratio $J_2/J_1$ there is 
an enhancement of $\kappa_{\rm nem}^{P}$ under the RG flow, and as one increases the ratio of $J_2/J_1$, the nematic response 
increases in a continuous manner up till $J_2/J_1=0.5$ where the system goes into collinear stripe antiferromagnetic order. However, 
throughout the paramagnetic regime, we do not observe any sudden pronounced increase of nematic responses that would indicate an 
onset of nematic order, and it is likely that the ground state found within PFFRG is symmetric.

\subsection{Comparison of VMC, DMRG, and PFFRG}
To summarize, the energy per site obtained by VMC and DMRG for the Hamiltonian~(\ref{eqn:heis-ham}) around $J_2/J_1 \approx 1/8$ 
are equal within error-bars. Nevertheless, while both VMC and DMRG agree on the fact that a magnetically disordered phase is
present close to the classical transition point, there are still discrepancies about the nature of this spin-liquid phase.
VMC finds an isotropic gapless $U(1)$ Dirac spin liquid as opposed to the DMRG results of a (potentially nematic) gapped 
$\mathbb{Z}_2$ spin liquid. As we add PFFRG to the picture in computing the spin susceptibility profile, we find that the
predominant short-range features such as peak enhancements at the corners of the Brillouin zone seem to be a common feature of 
nearly the whole paramagnetic domain around $J_2/J_1 \approx 1/8$ (see Fig.~\ref{fig:fig12}); however, from the equal-time spin
structure factor computed by VMC, this fact is shared by {\it both} the DSL and the symmetric $\mathbb{Z}_2\{\pi\}\mathcal{A}$ 
state [see Figs.~\ref{fig:fig10}(a) and~\ref{fig:fig10}(b)]. This short-range spin susceptibility picture is mainly confirmed by 
DMRG, whereas the nematic tendencies have not been found by VMC or PFFRG, and could possibly be rooted in the breaking of lattice 
point group symmetries from the outset, as found in the cylindrical geometries employed by DMRG. The latter cannot be the reason 
for the different findings from VMC and DMRG, as again, even allowing for nematic trial states in VMC still gives preference to 
the symmetric state.

How can different numerical methods find different ground states with equal energy per site and nearly identical short-range
spin susceptibility profiles? The controversy is likely to be rooted in the subtle interplay between short-range and long-range 
correlations, and how these are accounted for numerically. It is a well established feature that the long-wavelength behavior is 
difficult to pin down just by energetics, as, e.g., recently pointed out in Ref.~[\onlinecite{Balents-2014}]. 
This, in turn, is the only aspect by which the competing candidate states differentiate themselves: as bulk gaps become small the 
distinction between a gapless and a gapped state does become a matter of long-range correlations which might not be accurately taken 
into account from the viewpoint of energy densities. For the DSL predominantly found in VMC, not only for Eq.~(\ref{eqn:heis-ham}) 
but also for the Heisenberg model on the kagome lattice, one line of criticism comes from the fact that there are no rigorous proofs
that Dirac nodes in the fermion spectrum are stable in presence of $U(1)$ gauge fields (e.g., a monopole proliferation is expected
to generate a gapped phase). For the gapped spin liquids predominantly found in DMRG, there is threat of a bias toward lower entangled 
state, which likewise adds to the picture. In any case, because of the energy per site being even equal within error-bars for very 
different states, the paramagnetic regime of Eq.~(\ref{eqn:heis-ham}) is likely to become the paradigmatic test-bed for future 
investigations along these lines.

\section{Conclusions}\label{sec:conclusions}

In this work, we have shown that Gutzwiller-projected fermionic wave functions predicts a spin-liquid phase in the frustrated
Heisenberg model on the triangular lattice, for $0.08 \lesssim J_2/J_1 \lesssim 0.16$. This results is in excellent agreement with
previous DMRG calculations~\cite{White-2015,Hu-2015} (and also coupled-cluster results of Ref.~[\onlinecite{Li-2015}]), providing
an extremely competitive candidate to describe the paramagnetic ground state, namely the algebraic $U(1)$ Dirac spin liquid. This
fact contrasts the DMRG finding of a gapped (possibly nematic) $\mathbb{Z}_{2}$ spin liquid. The existence of a magnetically
disordered phase is also confirmed by PFFRG, which, however, cannot give definitive claims in the existence/absence of a spin gap.
Although the issue of the gap is very important and must be definitively settled down in the future, we would like to conclude by 
emphasizing the fact that, huge improvements of numerical techniques (including variational wave functions, DMRG approaches in
two spatial dimensions, and PFFRG for spin models) have been done in the recent past, allowing the community to reach considerably
important results in various contexts. The agreement on the ground-state energies in a wide parameter regime, but also on the extent
of the spin-liquid phase, represents a milestone from which other calculations will proceed. We are confident that, in the near
future, further developments of these approaches will give definitive answers to the unsolved problems.

{\it Acknowledgments}. We thank J. Reuther, S.-S. Gong, and S. Bieri for useful discussions, and R. Kaneko for providing us with the 
VMC data corresponding to Ref.~[\onlinecite{Kaneko-2014}]. R.T. thanks L.~Balents and S.~L.~Sondhi for insightful discussions. The 
work was supported by the European Research Council through ERC-StG-TOPOLECTRICS-Thomale-336012. Y.I. and R.T. thank the DFG (Deutsche 
Forschungsgemeinschaft) for financial support through SFB 1170. F.B. acknowledges support from Italian MIUR through PRIN 2010 
2010LLKJBX. D.P. acknowledges support from NQPTP Grant No. ANR-0406-01 from the French Research Council (ANR). W.-J.H. acknowledges 
support from the U.S. National Science Foundation through Grant No. DMR-1408560 (CSUN), Grant No. DMR-1350237 (Rice) and Grant No. DMR-1309531. D.P. and 
Y.I. acknowledge the CALMIP EOS cluster (Toulouse) for CPU time. Y.I. gratefully acknowledges the Gauss Centre for Supercomputing 
e.V. for funding this project by providing computing time on the GCS Supercomputer SuperMUC at Leibniz 
Supercomputing Centre (LRZ).


%

\end{document}